\documentclass[aps,pre,nofootinbib,amsmath,amsfonts]{revtex4}
\usepackage{hyperref}
\usepackage{graphicx, xcolor}
\usepackage{longtable}
\usepackage{subfigure}
\usepackage{euscript}

\newcommand{\eqr}[1]{Eq.~\eqref{#1}}

\newcommand{\secr}[1]{Sec.~[\ref{#1}]}
\newcommand{\figr}[1]{Fig.~[\ref{#1}]}
\newcommand{\ssecr}[1]{Subsec.~[\ref{#1}]}
\newcommand{\tblr}[1]{Table~[\ref{#1}]}
\newcommand{\tblsr}[2]{Tables~[\ref{#1}-\ref{#2}]}

\newcommand{\velocity}{{\rm{v}}}

\newcommand{\Kappa}{{\cal K}}

\newcommand{\alTcxi}{T, \, c, \, \zeta}
\newcommand{\alTxi}{T, \, \zeta}
\newcommand{\alTrhoxi}{T, \, \rho, \, \xi}

\newcommand{\alrhoxipr}{\rho, \, \xi, \, \rho', \, \xi'}
\newcommand{\alrhoxi}{\rho, \, \xi}

\newcommand{\vn}{{\mathbf{n}}}

\newcommand{\DX}[1]{\displaystyle{\frac{d #1}{dx}}}
\newcommand{\ddx}{\DX{}}
\newcommand{\dd}[1]{\displaystyle\frac{\partial}{\partial #1}}

\newcommand{\kr}[2]{\rho_{#1}\rho_{#2}}
\newcommand{\tsup}[1]{\textsuperscript{#1}}
\newcommand{\xs}{x^{\scriptstyle s}}
\newcommand{\xxs}{{\scriptstyle\{x^{\scriptstyle s}\}}}
\newcommand{\xgsb}{x^{g,s}_{\beta}}
\newcommand{\xlsb}{x^{\ell,s}_{\beta}}
\newcommand{\xg}{x^{g}}
\newcommand{\xl}{x^{\ell}}

\newcommand{\ssy}{\scriptstyle}

\newcommand{\profilescale}{0.4}
\newcommand{\figscale}{0.7}

\graphicspath{{Figures/}}

\begin{document}
\title{Numerical solution and verification of the local equilibrium for the flat interface in the two-phase binary mixture.}
\author{K.~S.~Glavatskiy}
\author{D.~Bedeaux}
\affiliation{Department of Chemistry, Norwegian University of Science and Technology, Trondheim, Norway}
\date\today
\begin{abstract}
In this paper we first apply the general analysis described in our first paper to a binary mixture of cyclohexane and $n$-hexane. We use the square gradient model for the continuous description
of a non-equilibrium surface and obtain numerical profiles of various thermodynamic quantities in various stationary state conditions. In the second part of this paper we focus on the
verification of local equilibrium of the surface as described with excess quantities. We give a definition of the temperature and chemical potential difference for the surface and verify that
these quantities are independent of the choice of the dividing surface. We verify that the non-equilibrium surface can be described in terms of Gibbs excess densities which are in good
approximation equal to their equilibrium values at the temperature and chemical potential difference of the surface.
\end{abstract}
\maketitle
%
%
\numberwithin{equation}{section}
\section{Introduction}
In a previous article \cite{glav/grad1}, referred to as paper I, we have established the general approach for the square gradient description of the interface between two phases in
non-equilibrium mixtures. We considered phenomena like temperature, density and mass fraction gradients; heat and diffusion fluxes as well as evaporation or condensation fluxes through the
interface. Some profiles were given, without going into details of the numerical procedures used to obtain them. In this paper we will do this.

In the general description of the interface one uses contributions to the Helmholtz free energy density proportional to the square of the density and mass fraction gradients. These contribution
imply that it is not possible to use \textit{continuous local equilibrium thermodynamics} in the interface, i.e. to calculate the local values of the various thermodynamic parameters in terms of
the local density, mass fractions and temperature only. Rowlinson and Widom (see \cite[page 43]{RowlinsonWidom}) use the name \textit{point thermodynamics} for this to distinguish it from other
\textit{quasi}- or \textit{local thermodynamic} treatments.  Given the non-autonomous nature of the square gradient model, it is sensible to question whether a description in terms of excess
variables along the lines given by Gibbs \cite{Gibbs/ScientificPapers}, can be autonomous. Gibbs' treatment, though only given for equilibrium systems, suggested such an assumption. This would
imply that the surface is a separate thermodynamic phase. Bakker \cite{Bakker/capilarity} and Guggenheim \cite[page 45]{Guggenheim/thermodynamics} made this assumption, the validity of which was
subsequently disputed by Defay and Prigogine \cite{DefayPrigogine/sta}. We refer to Rowlinson and Widom \cite[page 33]{RowlinsonWidom} for a discussion of this point. In the theory of
non-equilibrium thermodynamics of surfaces \cite{bedeaux/boundaryNE, bedeaux/advchemphys, albano/electrosurf, kjelstrupbedeaux/heterogeneous} Gibbs' description in terms of excess variables has
been used. It is then assumed that the Gibbs' description of the surface in terms of excess variables is autonomous, or in other words that one can use this property, which we will call
\textit{local equilibrium of the surface}, to describe the surface. In earlier work \cite{bedeaux/vdW/II} coauthored by one of us this property was verified for one-component systems. It is the
main objective of this paper to verify this property for binary mixtures. For details about the extension of the square gradient model to non-equilibrium systems we refer to paper I. For figures
of typical density, mass fraction and temperature profiles, and a discussion thereof, we also refer to the first paper. In this paper we focus on the properties of the excess variables.

We consider a flat interface between a binary liquid and it's vapor with the normal $\vn = (1,0,0)$ pointing from the vapor to the liquid. We assume that all fluxes and gradients gradients are
in the $x-$direction. Due to this all variables depend only on the $x-$coordinate. We assume the fluid to be non-viscous, so that the viscous pressure tensor $\pi _{\alpha \beta }=0$.

In \secr{sec/Equations} we will give all the equations, which are required for the determination of the profiles. We shall only consider stationary states in this paper. We choose the system
such that the gas is on the left hand side and the liquid is on the right hand side. The $x$-axes is directed from left to right and the gravitational acceleration $g$ is directed towards the
liquid. In \secr{sec/Procedure} we describe numerical procedure which have been used to solve these equations. We give the results in \secr{sec/Results/Profile}. We then proceed to the second
part of the article, the verification of local equilibrium of the surface. In \secr{sec/LocalEquilibrium} we introduce excesses and surface variables and discuss in general the meaning of local
equilibrium of the surface. In \secr{sec/Results/LocalEquilibrium} we give the results of the verification procedure. Finally, in \secr{sec/Conclusion} we give concluding remarks.

\section{Complete set of equations.}\label{sec/Equations}

In order to calculate the various profiles in the non-equilibrium mixture a complete set of equations is required. This set is given by the hydrodynamic balance (conservation) equations, the
thermodynamic equations of state and the phenomenological force-flux relations. These equations were given (derived) and discussed in paper I, to which we refer. In this paper we will only list
them for the stationary case.

\subsection{Conservation equations.}\label{sec/Equations/Conservation}

In a stationary state the conservation equations take the following form. The law of mass conservation is
\begin{equation}\label{eq/Equations/Conservation/01}
\begin{array}{rl}
\ddx\left(\rho\,\velocity\right) &= 0 \\
\\
\ddx\left(J_{1} + \rho\,\xi\,\velocity\right) &= 0
\end{array}
\end{equation}%
where $\rho =\rho_{1}+\rho_{2}$ and $\velocity=(\rho_{1}\,\velocity_{1}+\rho_{2}\,\velocity_{2})/\rho$ are the mass density and the barycentric velocity. Furthermore $\xi =\rho _{1}/\rho $ is
the mass fraction of the first component and $J_{1} = \rho\,\xi\,(\velocity_{1}-\velocity)$ is the diffusion flux of the first component relative to the barycentric frame of reference. Momentum
conservation is given by
\begin{equation}\label{eq/Equations/Conservation/02}
\begin{array}{l}
\ddx\left(\rho\,\velocity^{2} + p + \gamma_{xx}\right) = \rho\,g
\end{array}
\end{equation}%
where we call $\gamma _{\alpha \beta }$ the thermodynamic tension tensor, which will be defined below. Furthermore $p$ and $p+\gamma _{xx}$ are the pressures parallel and perpendicular to the
interface, for the planar interface under consideration. For curved surfaces see paper I. Energy conservation is given by
\begin{equation}\label{eq/Equations/Conservation/03}
\begin{array}{l}
\ddx\,J_{e} = 0
\end{array}
\end{equation}%
where $J_{e} \equiv J_{q} + \rho\,e\,\velocity + p\,\velocity$ is the total energy flux, $J_{q}$ is the heat flux, and $e = u + (\velocity^{2}/2 - g\,x)$ is the total specific energy and $u$ is
the specific internal energy.

\subsection{Thermodynamic equations.}\label{sec/Equations/Thermodynamic}

The gradient model, discussed in the first paper \cite{glav/grad1} gives the following expressions for the specific Helmholtz energy $f$, the specific internal energy $u$, the parallel pressure
$p$, the chemical potential difference $\psi \equiv \mu_{1}-\mu_{2}$, the chemical potential of the second component $\mu \equiv \mu_{2}$ and the $xx$-element of the tension tensor
$\gamma_{xx}$:
\begin{equation}\label{eq/Equations/Thermodynamic/01}
\begin{array}{rl}
f(x) &= f_{0}(\alTrhoxi) + \Kappa(\alrhoxipr) \\
\\
u(x) &= f_{0}(\alTrhoxi) - T\,\dd{T}f_{0}(\alTrhoxi) + \Kappa(\alrhoxipr) \\
\\
p(x)  &=  \rho^{2}\dd{\rho}\left(f_{0}(\alTrhoxi) + \Kappa(\alrhoxipr)\right) - \rho\,\ddx\left(\rho\,\dd{\rho'}\Kappa(\alrhoxipr)\right) \\
\\
\psi(x) & =  \dd{\xi}\left(f_{0}(\alTrhoxi) + \Kappa(\alrhoxipr)\right) - {\frac{1}{\rho}}\,\ddx\left(\rho\,\dd{\xi'}\Kappa(\alrhoxipr)\right) \\
\\
\mu(x) & =  \dd{\rho}\left(\rho\,\left(f_{0}(\alTrhoxi) + \Kappa(\alrhoxipr)\right)\right) - \psi(x)\,\xi - \rho\,\ddx\left(\rho\,\dd{\rho'}\Kappa(\alrhoxipr)\right) \\
\\
\gamma_{xx}(x) & =  2\,\rho\,\Kappa(\alrhoxipr)\\
\end{array}
\end{equation}
Here $f_{0}(\alTrhoxi) = f_{0}^{\nu}(\alTcxi)$ is the specific Helmholtz energy of the homogeneous phase, which can for instance be derived from the equation of state. Furthermore
$\Kappa(\alrhoxipr)$ is the gradient contribution, where the primes indicate a derivative with respect to $x$. Since the equation of state is usually given in molar quantities, it is convenient
to use them here as well. Thus, $c=\rho/M$ is the molar concentration, $\zeta =\xi\,M/M_{1}$ is the molar fraction, where $M=M_{1}\,M_{2}/(M_{1}+\xi\,(M_{2}-M_{1})) = M_{2}+\zeta\,(M_{1}-M_{2})$
is the molar mass of the mixture, and $M_{1}$ and $M_{2}$ are molar masses of each component.

\subsection{The homogeneous Helmholtz energy}

This energy is given by the following equation
\begin{equation}\label{eq/Equations/Thermodynamic/02}
f_{0}^{\nu}(\alTcxi) = - R\,T\,\ln\left({\frac{e}{c\,N_{A}}}{\frac{\mathrm{w}(\alTxi)}{\Lambda^{3}(\alTxi)}}\,\left(1-B(\zeta)\,c\right)\right) - A(\alTxi)\,c
\end{equation}
where the de Broglie wavelength $\Lambda$ and the characteristic sum over internal degrees of freedom w are respectively.

\begin{equation}  \label{eq/Equations/Thermodynamic/03}
\begin{array}{l}
\Lambda(\alTxi) = \hbar N_{A}\displaystyle\left(\frac{2\pi}{M R\,T}\right)^{1/2} \\
\\
\mathrm{w}(\alTxi) = \displaystyle\left(\frac{\mathrm{w}_{1}}{\zeta}\left({\frac{M_{1}}{M}}\right)^{\!3/2}\right)^{\zeta} \,
\left(\frac{\mathrm{w}_{2}}{1-\zeta}\left({\frac{M_{2}}{M}}\right)^{\!3/2}\right)^{1-\zeta}
\end{array}
\end{equation}
Expressions for the characteristic sums over internal degrees of freedom for each component, $\mathrm{w}_{1}$ and $\mathrm{w}_{2}$, are given in paper I. In this paper they are assumed to be
independent of the temperature and the molar fractions, i.e. just constant numbers.

The mixing rules for $A$ and $B$ are
\begin{equation}\label{eq/Equations/Thermodynamic/04}
\begin{array}{rcl}
A(\alTxi) &=&  a_{11}\,\zeta^{2} + 2\,a_{12}\,\zeta\,(1-\zeta) + a_{22}\,(1-\zeta)^{2}\\
B(\zeta) &=&  b_{1}\,\zeta + b_{2}\,(1-\zeta)
\end{array}
\end{equation}
with $a_{ij} = \sqrt{a_{i}\,a_{j}}$, where $a_{i}$ as well as $b_{i}$ is a coefficient of a pure component $i$. We will assume in this paper that $a_{ij}$ are independent of temperature.

\subsection{The gradient contribution}

This contribution is given by the following general expression for a binary mixture
\begin{equation}\label{eq/Equations/Thermodynamic/06}
\begin{array}{l}
\Kappa(\alrhoxipr) \equiv  \displaystyle{\frac{1}{2\,\rho}}\left(\kappa_{\rho\rho}(\alrhoxi)\,{\rho'}^{2} + 2\,\kappa_{\rho\xi}(\alrhoxi)\,{\rho'}\,{\xi'} + \kappa_{\xi\xi}(\alrhoxi)\,{\xi'}^{2}
\right)
\end{array}
\end{equation}
The coefficients $\kappa_{\rho\rho}$, $\kappa_{\rho\xi}$ and $\kappa_{\xi\xi}$ can be expressed in the gradient coefficients $\kappa_{\kr{1}{1}}$ and $\kappa_{\kr{2}{2}}$ for components 1 and 2
in the following way (see paper I for details)
\begin{equation}\label{eq/Equations/Thermodynamic/07}
\begin{array}{rcl}
\kappa_{\rho\rho}(\alrhoxi) &=& (\kappa_{\kr{1}{1}}-2\kappa_{\kr{1}{2}}+\kappa_{\kr{2}{2}})\,\xi^{2} + 2\,(\kappa_{\kr{1}{2}}-\kappa_{\kr{2}{2}})\,\xi + \kappa_{\kr{2}{2}}
\\
\kappa_{\rho\xi}(\alrhoxi) &=& (\kappa_{\kr{1}{1}}-2\kappa_{\kr{1}{2}}+\kappa_{\kr{2}{2}})\,\rho\,\xi + (\kappa_{\kr{1}{2}}-\kappa_{\kr{2}{2}})\,\rho
\\
\kappa_{\xi\xi}(\alrhoxi) &=& (\kappa_{\kr{1}{1}}-2\kappa_{\kr{1}{2}}+\kappa_{\kr{2}{2}})\,\rho^{2}
\end{array}
\end{equation}
where we use the mixing rule $\kappa_{\kr{1}{2}} = \sqrt{\kappa_{\kr{1}{1}}\,\kappa_{\kr{2}{2}}}$ similar to the one for coefficients $a_{ij}$ for the cross coefficient. We will assume
$\kappa_{\kr{i}{j}}$ to be independent of the densities in this paper.

With the above mixing rules the gradient contribution can be written in the form
\begin{equation}\label{eq/Equations/Thermodynamic/08}
\begin{array}{l}
\Kappa(\alrhoxipr) \equiv  \displaystyle{\frac{\quad\kappa\,{q'}^{2}}{2\,\rho}}
\end{array}
\end{equation}
where $\kappa \equiv \kappa_{\kr{2}{2}}$ and $q \equiv \rho\,(1+\varepsilon_{\kappa}^{m}\,\xi)$, where $\varepsilon_{\kappa}^{m} \equiv \varepsilon_{\kappa} \equiv
(\sqrt{\kappa_{\kr{1}{1}}}-\sqrt{\kappa_{\kr{2}{2}}})/\sqrt{\kappa_{\kr{2}{2}}}$. Some of the quantities from \eqr{eq/Equations/Thermodynamic/01} can be rewritten as
\begin{equation}\label{eq/Equations/Thermodynamic/09}
\begin{array}{rl}
p(x)  &=  p_{0} - \kappa\,\left({1 \over 2}\,{q'}^{2} + q\,q''\right)\\
\\
\mu(x) & =  \mu_{0} - \kappa\,q''\\
\\
\psi(x) & =  \psi_{0} - \varepsilon_{\kappa}\,\kappa\,q''\\
\end{array}
\end{equation}
where $p_{0}$, $\mu _{0}$ and $\psi _{0}$ are values of the corresponding quantities in the homogeneous phase, which are found from \eqr{eq/Equations/Thermodynamic/01} by setting $\Kappa = 0$.
For a one-component fluid $q$ equals the density. For the two-component mixture $q$ plays a similar role as the density for the one-component fluid. We shall therefore refer to $q$ as the order
parameter.

The size of the coefficient $\varepsilon_{\kappa}$ depends on the nature of the components of the mixture. In an organic mixture like cyclohexane and n-hexane, a mixture we will study in more
detail in this paper, the components are very similar and as a consequence $\varepsilon_{\kappa}$ is small. The order parameter is then in good approximation equal to the density. When the
components are very different $\left\vert\varepsilon_{\kappa }\right\vert$ may be large and $q$ may become in good approximation equal to the density of one of the components (for instance, the
surface tensions of acetone and carbon disulfide are 22.7 and 31.3 gs\tsup{-2}. Thus the values of corresponding $\kappa_{\kr{i}{i}}$ are not very close and in their mixture the value of
$\varepsilon_{\kappa}$ may be compared to 1).

\subsection{Phenomenological equations.}\label{sec/Equations/Phenomenological}

In paper I we derived the general expression for the entropy production of a mixture in the interfacial region. For a non-viscous binary mixture which has only gradients and fluxes in the
$x-$direction it takes the following form
\begin{equation}\label{eq/Equations/Phenomenological/01}
\sigma_{s} = J_{q}\,\ddx{\frac{1}{T}} - J_{1}\,\ddx\frac{\psi}{T}
\end{equation}
The resulting linear force-flux relations are:
\begin{equation}\label{eq/Equations/Phenomenological/02}
\begin{array}{rl}
\ddx{\frac{1}{T}}       &= R_{qq}\,J_{q} - R_{q1}J_{1}\\
\\
\ddx{\frac{\psi}{T}}    &= R_{1q}\,J_{q} - R_{11}J_{1}\\
\end{array}
\end{equation}
The resistivity coefficients $R_{qq}$, $R_{11}$ and $R_{q1}=R_{1q}$ will in general depend on the densities, their gradients as well as on the temperature, so they vary through the interface.
Expressions for the resistivity profiles in the interfacial region are not available. We model them, using the bulk values as the limiting value away from the surface and the order parameter
profile as a modulatory curve.
\begin{equation}\label{eq/Equations/Phenomenological/03}
\begin{array}{rl}
R_{qq}(x) &= R_{qq}^{g} + (R_{qq}^{\ell}-R_{qq}^{g})\,q_{0}(x) + \alpha_{qq}(R_{qq}^{\ell}+R_{qq}^{g})\,q_{1}(x)\\
\\
R_{q1}(x) &= R_{q1}^{g} + (R_{q1}^{\ell}-R_{q1}^{g})\,q_{0}(x) + \alpha_{q1}(R_{q1}^{\ell}+R_{q1}^{g})\,q_{1}(x)\\
\\
R_{11}(x) &= R_{11}^{g} + (R_{11}^{\ell}-R_{11}^{g})\,q_{0}(x) + \alpha_{11}(R_{11}^{\ell}+R_{11}^{g})\,q_{1}(x)\\
\end{array}
\end{equation}
where
\begin{equation}\label{eq/Equations/Phenomenological/04}
q_{0}(x) = {\frac{q(x)-q^{g}_{eq}}{q^{\ell}_{eq}-q^{g}_{eq}}}, \qquad\qquad  q_{1}(x) = \frac{|q'(x)|^{2}}{|q'_{eq}(x)|_{max}^{2}}
\end{equation}
are modulatory curves for resistivity profiles. Here $q_{eq}^{g}$ and $q_{eq}^{\ell }$ are the equilibrium coexistence values of the order parameter of the gas and liquid respectively.
Furthermore $|q_{eq}^{\prime}(x)|_{max}^{2}$ is the maximum value of the squared equilibrium order parameter gradient. For each resistivity profile $R^{g}$ and $R^{\ell}$ are the equilibrium
coexistence resistivities of the gas and liquid phase respectively. Coefficients $\alpha_{qq}$, $\alpha_{q1}$, $\alpha_{11}$ control the size of the gradient term, which gives peaks in the
resistivity profiles in the interfacial region. Such a peak is observed in molecular dynamic simulations of one -component fluids \cite{surfres}.

Limiting coefficients $R^{b}$ (where $b$ is either $g$ or $\ell$) are related to measurable transport coefficients in the bulk phases: thermal conductivity $\lambda^{b}$, diffusion coefficient
$D^{b}$ and Soret coefficient $s_{T}^{b}$. In the description of transport in the \textit{bulk} phases it is convenient to use measurable heat fluxes
\begin{equation}\label{eq/Equations/Phenomenological/05a}
J_{q}^{\prime\,b} = J_{q}^{b} - J_{1}^{b}(h_{1}^{b} - h_{2}^{b})
\end{equation}
where $h_{i}^{b}$ is a specific enthalpy of component $i$ in phase $b$. Furthermore we used that $J_{2}^{b} = -J_{1}^{b}$. In the bulk phases the entropy production then takes the following
form:
%
\begin{equation}\label{eq/Equations/Phenomenological/05}
\sigma_{s} = J_{q}'\,\ddx{\frac{1}{T}} - J_{1}\,{\frac{1}{T}}\DX{\psi_{T}}
\end{equation}
where we have suppressed the superscript $b$ for now. The subscript $T$ of $\psi$ indicates that the gradient is calculated keeping the temperature constant. Using Gibbs-Duhem relation in a
homogeneous phase at a constant pressure one can show, that
\begin{equation}\label{eq/Equations/Phenomenological/06a}
\DX{\psi_{T}} = {\frac{\partial\mu_{1}}{\partial \xi}}{\frac{1}{(1-\xi)}}\DX{\xi}
\end{equation}
After introducing measurable transport coefficients, the force-flux relations derived from \eqr{eq/Equations/Phenomenological/05} can be written in a form used in \cite{deGrootMazur} :
\begin{equation}\label{eq/Equations/Phenomenological/06}
\begin{array}{rcl}
J_{q}'&=& -\lambda\,\DX{T} - \rho\,\xi\,{\frac{\partial\mu_{1}}{\partial \xi}}\,T\,D\,s_{T}\,\DX{\xi}\\
\\
J_{1} &=& -\rho\,\xi\,(1-\xi)\,D\,s_{T}\,\DX{T} - \rho\,D\,\DX{\xi}\\
\end{array}
\end{equation}
Comparing \eqr{eq/Equations/Phenomenological/02} and \eqr{eq/Equations/Phenomenological/06} in the bulk region we find the bulk values $R$ of corresponding resistivity coefficients
\begin{equation}\label{eq/Equations/Phenomenological/07}
\begin{array}{rl}
R_{qq} &= \displaystyle{\frac{1}{L\,T^{2}}}{\frac{D\,\rho}{\psi_{\xi}}}\\
\\
R_{q1} = R_{1q} &= \displaystyle{\frac{1}{L\,T^{2}}}\left({\frac{D\,\rho}{\psi_{\xi}}}\,(h_{1}-h_{2}) + D\,s_{T}\,\rho\,\xi\,(1-\xi)\,T\right)\\
\\
R_{11} &= \displaystyle{\frac{1}{L\,T^{2}}}\left({\frac{D\,\rho}{\psi_{\xi}}}\,(h_{1}-h_{2})^{2} + D\,s_{T}\,\rho\,\xi\,(1-\xi)\,T\,(h_{1}-h_{2}) + \lambda\,T \right)\\
\end{array}
\end{equation}
where $\psi_{\xi} = (\partial\psi/\partial\xi)$, $L = (\lambda\,D\,\rho/\psi_{\xi}) - (D\,s_{T}\,\rho\,\xi\,(1-\xi))^{2}\,T$. All the quantities in \eqr{eq/Equations/Phenomenological/07} are
taken in the specified bulk, either gas or liquid.

\section{Solution procedure.}\label{sec/Procedure}

The numerical procedure is similar to the one, described in \cite{bedeaux/vdW/I}, however it has some differences. We will describe the special features below. We use the Matlab procedure
\texttt{bvp4c} \cite{bvp4c} to solve the stationary boundary value problem. It requires a reasonable initial guess and boundary conditions. We use the equilibrium profile as the initial guess.
We use a box of width 80 nm with the grid containing of 81 equidistantly spread points.

\subsection{Equilibrium profile.}\label{sec/Procedure/Equilibrium}

It is easier to describe equilibrium properties of the mixture using molar quantities. Everywhere in this subsection we will do this. The superscript $\nu$ indicates a molar quantity. The total
molar concentration and molar fraction of the first component are denoted by $c$ and $\zeta$ respectively.

Equilibrium coexistence is determined by the following system of equations
\begin{equation}\label{eq/Procedure/Equilibrium/01}
\begin{array}{rcl}
\mu_{eq}^{\nu} =& \mu_{0}^{\nu}(T_{eq}, c^{g}_{eq}, \zeta^{g}_{eq}) &= \mu_{0}^{\nu}(T_{eq}, c^{\ell}_{eq}, \zeta^{\ell}_{eq})\\
\psi_{eq}^{\nu} =& \psi_{0}^{\nu}(T_{eq}, c^{g}_{eq}, \zeta^{g}_{eq}) &=  \psi_{0}^{\nu}(T_{eq}, c^{\ell}_{eq}, \zeta^{\ell}_{eq})\\
p_{eq} =& p_{0}(T_{eq}, c^{g}_{eq}, \zeta^{g}_{eq}) &=  p_{0}(T_{eq}, c^{\ell}_{eq}, \zeta^{\ell}_{eq})\\
\end{array}
\end{equation}
where $\psi_{0}^{\nu} = (\partial f_{0}^{\nu} / \partial \zeta)$, $\mu_{0}^{\nu} = f_{0}^{\nu} + c\,(\partial f_{0}^{\nu} / \partial c) - \psi_{0}^{\nu}\,\zeta $ and $p_{0} = c^{2}(\partial
f_{0}^{\nu} / \partial c)$ are homogeneous chemical potentials and pressure. $c^{g}_{eq}$, $\zeta^{g_{eq}}$ and $c^{\ell}_{eq}$, $\zeta^{\ell}_{eq}$ are coexistence density and mass fractions of
gas and liquid respectively.

Having 6 equations \eqr{eq/Procedure/Equilibrium/01} and 8 unknowns $c^{g}_{eq}$, $\zeta^{g}_{eq}$, $c^{\ell}_{eq}$, $\zeta^{\ell}_{eq}$ and $\psi_{eq}^{\nu}$, $\mu_{eq}^{\nu}$, $p_{eq}$,
$T_{eq}$, an equilibrium state for two-phase two component mixture contains two free parameters. Particularly, the temperature and the molar fraction of the liquid phase are experimentally a
reasonable choice. We have found, however, that it is more convenient to control $T_{eq}$ and $\psi_{eq}^{\nu}$ in the calculations. $\psi_{eq}^{\nu}$ changes monotonically with $\zeta^{g}_{eq}$
or $\zeta^{\ell}_{eq}$, and it is therefore a good measure for the composition. Since $\psi^{\nu} = \mu_{1}^{\nu} - \mu_{2}^{\nu}$, the value of $\psi^{\nu}$ gives the difference of the chemical
potentials of two components.

To obtain the equilibrium profiles $c_{eq}(x)$ and $\zeta_{eq}(x)$ one needs to solve a system of two differential equations
\begin{equation}\label{eq/Procedure/Equilibrium/02}
\begin{array}{rl}
\mu_{eq}^{\nu} & =  \mu_{0}^{\nu}(c, \zeta) - \kappa^{\nu}\,(q^{\nu})''\\
\\
\psi_{eq}^{\nu} & =  \psi_{0}^{\nu}(c, \zeta) - \varepsilon_{\kappa}^{\nu}\,\kappa^{\nu}\,(q^{\nu})''\\
\end{array}
\end{equation}
where $q^{\nu} = c\,(1+\varepsilon_{\kappa}^{\nu}\,\zeta)= q\,M_{2}$, $\kappa^{\nu} = \kappa\,M_{2}^{2}$ and $\varepsilon_{\kappa}^{\nu} = (1+\varepsilon_{\kappa})\,(M1/M2)-1$ where $M_{1}$ and
$M_{2}$ are the molar masses of the components. This system of equations is, in fact, singular, since coefficients of the higher derivatives are proportional. Thus, we can derive one algebraic
equation instead of a differential one.

\begin{equation}\label{eq/Procedure/Equilibrium/03}
\psi_{0}^{\nu}(c, \zeta) - \varepsilon_{\kappa}^{\nu}\,\mu_{0}^{\nu}(c, \zeta) =  \psi_{eq}^{\nu} - \varepsilon_{\kappa}^{\nu}\,\mu_{eq}^{\nu}\\
\end{equation}
The \texttt{bvp4c} procedure takes only differential equations, so we have to transform \eqr{eq/Procedure/Equilibrium/03} to a differential one. This can be done easily by taking derivative of
both sides. After some transformations, we have the following equation set
\begin{equation}\label{eq/Procedure/Equilibrium/04}
\begin{array}{rl}
(q^{\nu})'' & =  \displaystyle{1 \over \kappa^{\nu}}\,\left(\mu_{0}^{\nu}(c, \zeta) - \mu_{eq}^{\nu}\right)\\
\\
\zeta' &= -(q^{\nu})'\, \displaystyle\left((1+\varepsilon_{\kappa}^{\nu}\,\zeta)\,\frac{\psi^{\nu}_{0\,\zeta}-\varepsilon_{\kappa}^{\nu}\,\mu^{\nu}_{0\,\zeta}}{\psi^{\nu}_{0\,c}-\varepsilon_{\kappa}^{\nu}\,\mu^{\nu}_{0\,c}} - \frac{\varepsilon_{\kappa}^{\nu}\,q}{1+\varepsilon_{\kappa}^{\nu}\,\zeta}  \right)^{-1} \\
\end{array}
\end{equation}
where subscripts $\zeta$ or $c$ mean partial derivative of the corresponding quantity with respect to $\zeta$ or $c$. This is the system of 3 first order differential equations for 3 variables
$\zeta$, $q^{\nu}$ and $(q^{\nu})'$, which requires 3 boundary conditions. One of them is \eqr{eq/Procedure/Equilibrium/03} taken on one of the boundaries, which simply determines the
integration constant for the second differential equation. The other two are $(q^{\nu})'(x_{g}) = 0$ and $(q^{\nu})'(x_{\ell}) = 0$ which indicate the fact, that box boundaries are in the
homogeneous region.

Numerical procedure allows any values of the variables. However, not all the values are allowed physically. For instance, mol fraction $\zeta$ is bounded in the interval $(0; 1)$ and molar
concentration is bounded in the interval $(0; B^{-1})$, where $B$ is given in \eqr{eq/Equations/Thermodynamic/04}. In order to avoid out-of-range problems, we use the function which safely maps
unit interval to real axes and vice versa:
\begin{equation}\label{eq/Procedure/Equilibrium/05}
\text{u2r}(u) = \arcsin(2u-1) \qquad\text{and}\qquad \text{r2u}(r) = 0.5(1+\sin(r))
\end{equation}
Particularly, the actual variables, we provide to \texttt{bvp4c} procedure are
\begin{equation}\label{eq/Procedure/Equilibrium/06}
\begin{array}{rl}
Y_{1} &= \text{u2r}(\zeta)\\
\\
Y_{2} &= \text{u2r}(q^{\nu}/q^{\nu}_{\infty})\\
\\
Y_{3} &= \text{u2r}\,'(q^{\nu}/q^{\nu}_{\infty},\, (q^{\nu})'/q^{\nu}_{\infty})
\end{array}
\end{equation}
where $\text{u2r}\,'(u, u') = u'/\sqrt{u - u^{2}}$ is simply the derivative of $\text{u2r}$ and $q^{\nu}_{\infty} = B^{-1}\max(1, |\varepsilon_{\kappa}^{\nu}|)$ is the limiting value for
$q^{\nu}$.

\subsection{Non-equilibrium profile.}\label{sec/Procedure/NonEquilibrium}

Non-equilibrium conditions are implemented when we change temperature or pressure from their equilibrium values. This results in mass and heat fluxes through the interface. The amount of matter
will then change in the gas and liquid phase. We will put the system in such conditions, that the total contents of the box is constant and equal to the equilibrium contents. It means, that if
some amount of liquid has been evaporated, the same amount of gas is condensed externally and put back into the liquid phase.

We introduce the overall mass $m(x) = \int_{x_{g}}^{x}{dy\,\rho(y)}$ and the mass of the 1st component $m_{\xi}(x) = \int_{x_{g}}^{x}{dy\,\rho(y)\,\xi(y)}$, which obey the following equations by
definition
\begin{equation}\label{eq/Procedure/NonEquilibrium/01}
\begin{array}{rl}
m'(x) &= \rho(x) \\
m_{\xi}'(x) &= \rho(x)\,\xi(x)
\end{array}
\end{equation}

We introduce the the overall mass flux $J_{m}$, the mass flux of the 1st component $J_{\xi}$, the energy flux $J_{e}$ and the "pressure" flux $J_{p}$:
\begin{equation}\label{eq/Procedure/NonEquilibrium/02}
\begin{array}{rl}
J_{m} &=\displaystyle \rho\,\velocity \\ \\
J_{\xi} &=\displaystyle J_{1} + \xi\,J_{m} \\
J_{e} &=\displaystyle  J_{q} + J_{m}\,\left(u_{0} + {1 \over 2}\left({J_{m} \over \rho}\right)^{2} + {1 \over \rho}\left(p_{\perp} - {1 \over 2}\kappa\,{q'}^{2}\right) - g\,x \right)\\
J_{p} &=\displaystyle  p_{\perp} + \frac{J_{m}^{2}}{\rho} - m\,g\\
\end{array}
\end{equation}
From \ssecr{sec/Equations/Conservation} one can see, that all these fluxes are constant.

From \eqr{eq/Equations/Thermodynamic/09} we obtain
\begin{equation}\label{eq/Procedure/NonEquilibrium/03}
\begin{array}{rl}
\kappa\,q'' &=  \displaystyle{1 \over q} \left(p_{0} - p_{\perp} + {1 \over 2}\,{q'}^{2} \right) \\
\\
\kappa\,q'' &=  \displaystyle{1 \over \varepsilon_{\kappa}} \left(\psi_{0} -  \psi \right)\\
\end{array}
\end{equation}
since $p_{\perp}(x) = p(x) + \kappa\,{q'}^{2}$. As in \eqr{eq/Procedure/Equilibrium/02} we have a singular set which leads to the algebraic equation
\begin{equation}\label{eq/Procedure/NonEquilibrium/04}
\psi_{0} - \psi - {\varepsilon_{\kappa} \over q} \left(p_{0} - p_{\perp} + {1 \over 2}\,{q'}^{2}\right)  = 0
\end{equation}
Taking derivative of this equation with respect to coordinate we obtain the expression for the first derivative of the fraction (we do not give the exact expression, since it's very complicated)
\begin{equation}\label{eq/Procedure/NonEquilibrium/05}
\xi' = \xi'(q, q', \xi, T, \psi, J_{m}, J_{\xi}, J_{e}, J_{p})
\end{equation}

As a consequence we have 7 unknown variables, $q$, $q'$, $m$, $m_{\xi}$, $\xi$, $T$, $\psi$ and 4 unknown fluxes $J_{m}$, $J_{\xi}$, $J_{e}$, $J_{p}$. This requires 7 first order differential
equations and 11 boundary conditions (7 of them determine integration constants of differential equations and 4 of them determine constant fluxes). As differential equations we use
\eqr{eq/Procedure/NonEquilibrium/01}, \eqr{eq/Equations/Phenomenological/02}, one of \eqr{eq/Procedure/NonEquilibrium/03} and \eqr{eq/Procedure/NonEquilibrium/05}. As boundary conditions we use
the following: The first boundary condition is \eqr{eq/Procedure/NonEquilibrium/04} taken on one of the boundaries, which simply determine the integration constant for
\eqr{eq/Procedure/NonEquilibrium/05}. The 4 other conditions control the overall content of the box (particularly, it is the same as in equilibrium)
\begin{equation}\label{eq/Procedure/NonEquilibrium/06}
\begin{array}{rl}
m(x_{g}) &= 0 \\
m_{\xi}(x_{g}) &= 0\\
m(x_{\ell}) &= m_{eq} \\
m_{\xi}(x_{\ell}) &= m_{\xi,\,eq}\\
\end{array}
\end{equation}
Here $m_{eq}$ and $m_{\xi,\,eq}$ are the equilibrium values of the total overall mass and the total mass of the 1st component in the whole box. The 2 more conditions are
\begin{equation}\label{eq/Procedure/NonEquilibrium/07}
\begin{array}{rl}
q'(x_{g}) &= 0 \\
q'(x_{\ell}) &= 0 \\
\end{array}
\end{equation}
which indicate the fact, that box boundaries are in the homogeneous region. In contrast to the equilibrium case, the density in the non-equilibrium homogeneous region may vary with coordinate,
so $q'$ differs from zero on the boundaries, and, in fact, they do. The value of the $q'$ is however small in the homogeneous region, comparing to the value in the surface region, so we may
neglect it and use such approximation. This will lead to the wrong profile behavior only in the small vicinity on the boundary, which we will exclude from the further analysis. The 4 last
boundary conditions reflect the conditions of the probable experiment. For instance, we may control the temperatures on the both side of the box, the pressure of the vapor side and the fraction
on the liquid side.
\begin{equation}\label{eq/Procedure/NonEquilibrium/08}
\begin{array}{rl}
T(x_{g}) &= T^{g} \\
T(x_{\ell}) &= T^{\ell} \\
p_{\perp}(x_{g}) &=p^{g} \\
\xi(x_{\ell}) &= \xi^{\ell} \\
\end{array}
\end{equation}

To solve these equations numerically we use the same technics as in the equilibrium case. All the variables should be properly scaled in order to make them to be the same order of magnitude.
This balances the numerical residual and gives better solution result. We use the following variables
\begin{equation}\label{eq/Procedure/NonEquilibrium/09}
\begin{array}{rl}
Y_{1} &= \text{u2r}(\xi)\\
\\
Y_{2} &= \text{u2r}(q/q_{\infty})\\
\\
Y_{3} &= \text{u2r}\,'(q/q_{\infty},\, q'/q_{\infty})\\
\\
Y_{4} &= m/(x^{*}q^{*})\\
\\
Y_{5} &= m_{\xi}/(x^{*}q^{*})\\
\\
Y_{6} &= T^{*}/T\\
\\
Y_{7} &= (\psi/T)(T^{*}/\psi^{*})\\
\end{array}
\end{equation}
where $q_{\infty} = q^{\nu}_{\infty}/M_{2}$ and scaling parameters $x^{*} = x_{\ell}-x_{g}$, $T^{*} = T_{eq}$, $\psi^{*} = \psi_{eq}$, $q^{*} = p_{eq}/\psi_{eq}$.

\section{Results for the temperature and chemical potential profiles.}\label{sec/Results/Profile}

In this section we show some profiles, obtained with the help of the above procedure. We choose a mixture of hexane and cyclohexane and give some of their properties relevant for our calculation
in the table below. We note, that among them only the molar masses have been measured. There are number of problems to obtain the values of other material properties. We determined, for
instance, the two van der Waals coefficients of the pure phases using their critical temperatures and pressures\footnote{In this we follow the example of the Handbook of Chemistry and Physics
\cite{Data/HBCP} rather than refs. \cite{bedeaux/vdW/I, bedeaux/vdW/II} where $T_{c}$\ and $v_{c}$\ were used.}. As a consequence the critical volumes per mole found in our description for the
pure components differs substantially from the experimental value. For the mixture the van der Waals coefficients were then found using the mixing rules. We use the values of the molar mass and
the van der Waals coefficients given in \tblr{tbl/const/material}:
\begin{center}\begin{longtable}{l @{\;\qquad} l @{\;\qquad} l  @{\;\qquad} l @{\;\qquad}}
\caption{The molar mass and the van der Waals coefficients} \label{tbl/const/material} \\
\hline
component & $M$, $\times 10^{-3}$ kg/mol & $a$, J m\tsup{3}/mol\tsup{2}  & $b$, $\times 10^{-5}$ m\tsup{3}/mol  \\
\hline
1 & 84.162              & 2.195                         & 14.13                         \\
2 & 86.178              & 2.495                         & 17.52                         \\
\hline
\end{longtable}%
\end{center}

Transfer coefficients of homogeneous fluids depend on temperature and densities, while these dependencies are not always available. We use typical constant values of these coefficients at the
conditions, close to the above equilibrium conditions. The values of the heat conductivity $\lambda$ are well-tabulated and we take them from \cite{Data/Transport/Yaws}. We take the typical
value of the diffusion coefficient $D$ for a liquid mixture from \cite{Data/Transport/Dong} and use the argument from \cite[p.279]{deGrootMazur} to obtain the typical value of the diffusion
coefficient for a gas mixture. Another argument from \cite[p.279]{deGrootMazur} is used to get the typical value of the Soret coefficient $s_{T}$. We use the data given in
\tblr{tbl/const/transfer}
\begin{center}%
\begin{longtable}{l @{\;\qquad} l @{\;\qquad\qquad} l @{\;\qquad} l @{\;\qquad} l }
\caption{Transfer coefficients} \label{tbl/const/transfer} \\
\hline
                & \multicolumn{2}{c}{$\lambda$, W/(m K)}    & $D$, m\tsup{2}/s  & $s_{T}$, 1/K   \\
\cline{2-3}
phase $\backslash$ component                & 1        & 2                              &                   &               \\
\hline
gas        & 0.0140  & 0.0157                          & 3.876 x $10^{-5}$ &    $10^{-4}$  \\
liquid  & 0.1130  & 0.1090                          & 3.876 x $10^{-9}$ &    $10^{-4}$  \\
\hline
\end{longtable}%
\end{center}
together with the "mixing" rules for the heat conductivity
\begin{equation}\label{eq/Results/Profile/01}
\begin{array}{rl}
\lambda^{g} &= \xi^{q}_{eq}\,\lambda^{g}_{1} + (1-\xi^{q}_{eq})\,\lambda^{g}_{2} \\
\lambda^{\ell} &= \xi^{\ell}_{eq}\,\lambda^{\ell}_{1} + (1-\xi^{\ell}_{eq})\,\lambda^{\ell}_{2} \\
\end{array}
\end{equation}

The values of the gradient coefficients are not available at all. One can determine them comparing the actual value of the surface tension of a pure fluid with the one, calculated with a given
$\kappa_{\kr{i}{j}}$. For given conditions the value of the surface tension of the mixture is about 0.027 N/m. We therefore choose $\kappa^{\nu}$ to be equal $12 \times 10^{-18}$ J
m\tsup{5}/mol\tsup{2} and $\varepsilon_{\kappa}^{\nu} = 0.01$. This gives values of the surface tension about 0.03 N/m.

The equilibrium properties of the system are calculated\footnote{From now on we will use specific quantities per unit of mole in the description. We will omit the superscript $\nu$ in the
following sections.} at $T_{eq}=330$ K and $\psi^{\nu}_{eq} = 700$ J/mol. This gives $p_{eq} = 376095$~Pa, $\mu^{\nu}_{eq} = -57098$~J/mol, $c^{g}_{eq} = 153.23$~mol/m\tsup{3}, $c^{\ell}_{eq} =
4898.26$~mol/m\tsup{3}, $\zeta^{g}_{eq} = 0.5519$ and $\zeta^{\ell}_{eq} = 0.5934$.

\begin{figure}
\centering
\subfigure[Temperature ] %
{\includegraphics[scale=\profilescale]{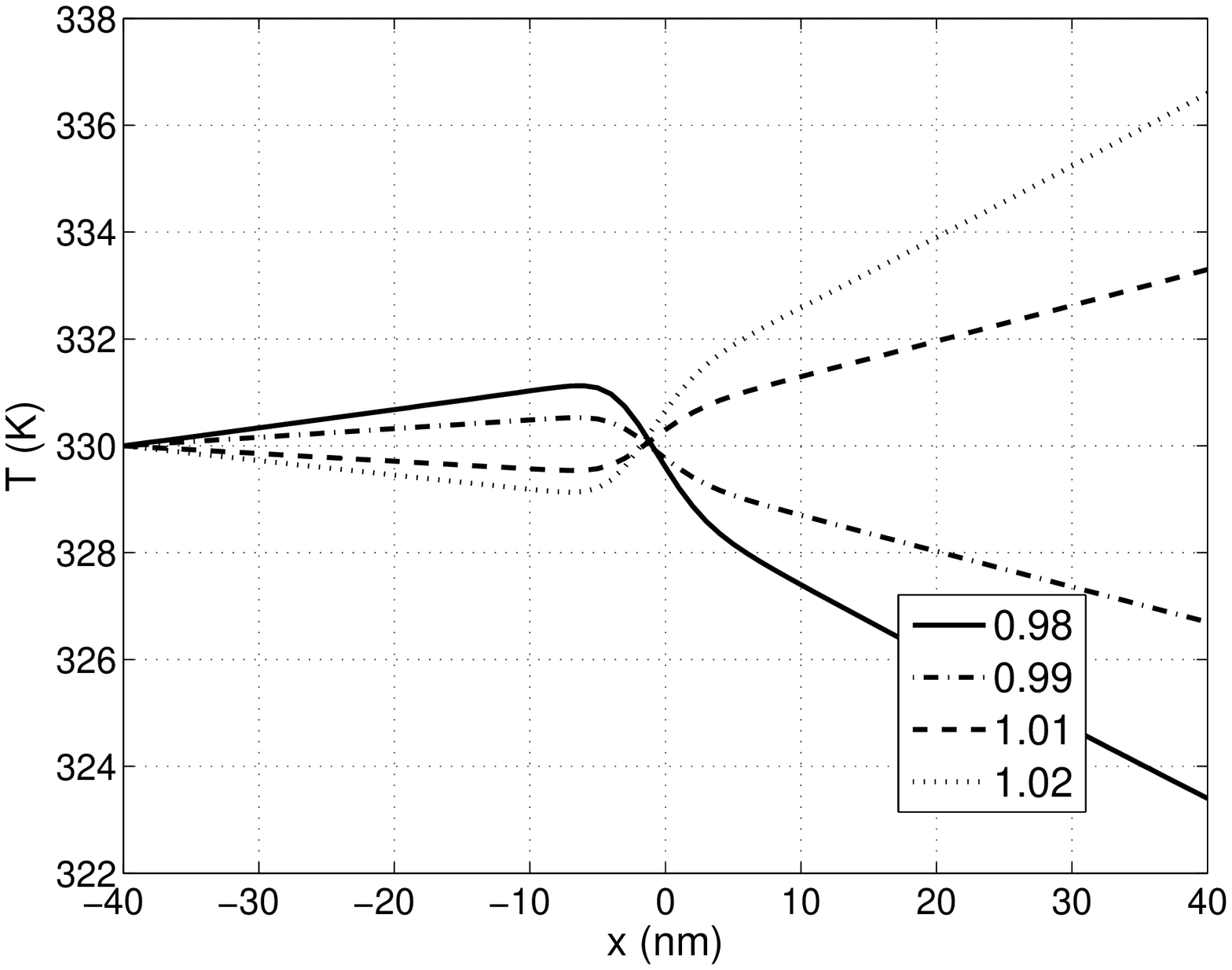}\label{Fig_T_Tl} } %
\subfigure[Chemical potential ] %
{\includegraphics[scale=\profilescale]{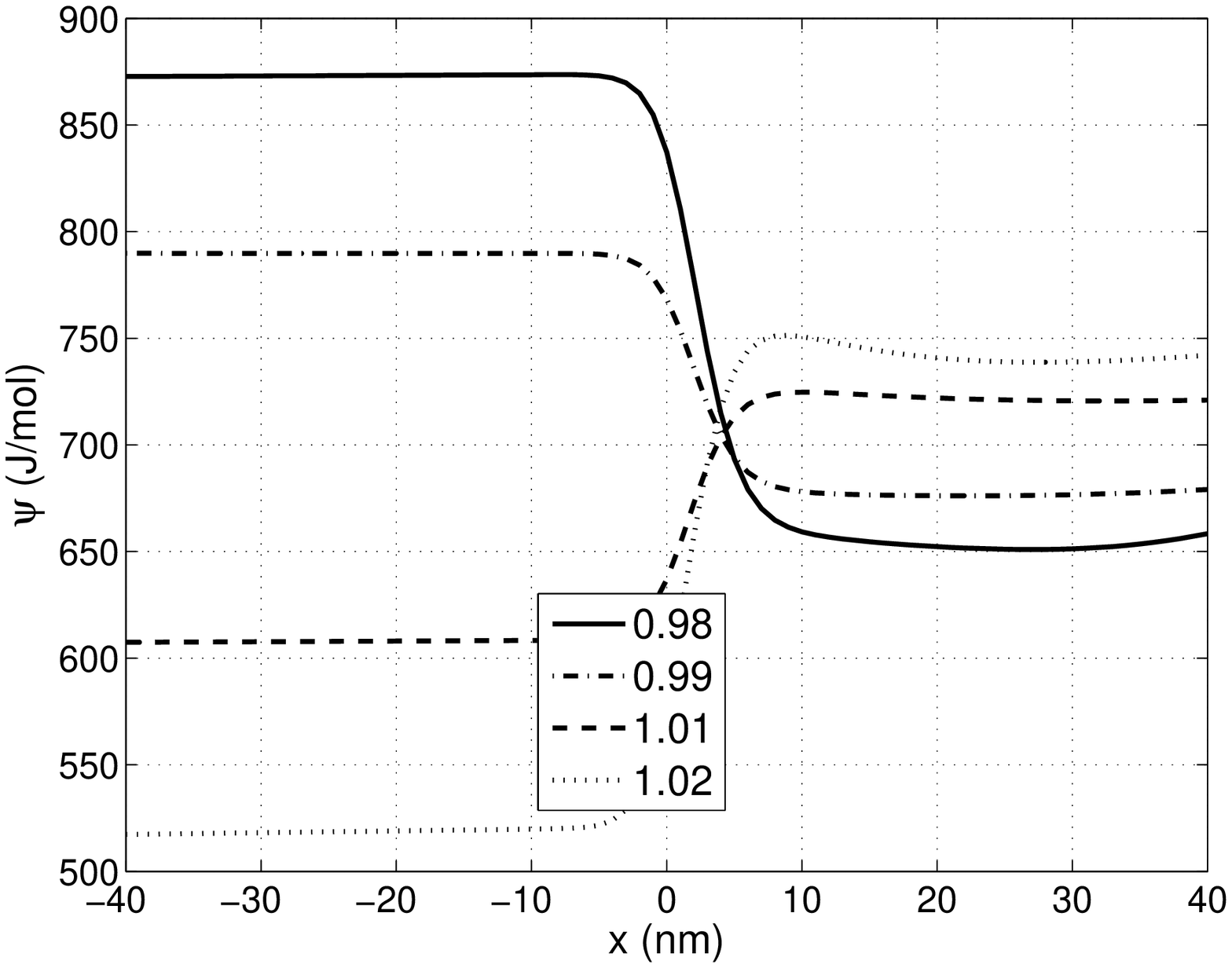} \label{Fig_psi_Tl} } %
\caption{Temperature and chemical potential profiles at various $T^{\ell}$}\label{Fig_Tl}
\end{figure}

\begin{figure}
\centering
\subfigure[Temperature ] %
{\includegraphics[scale=\profilescale]{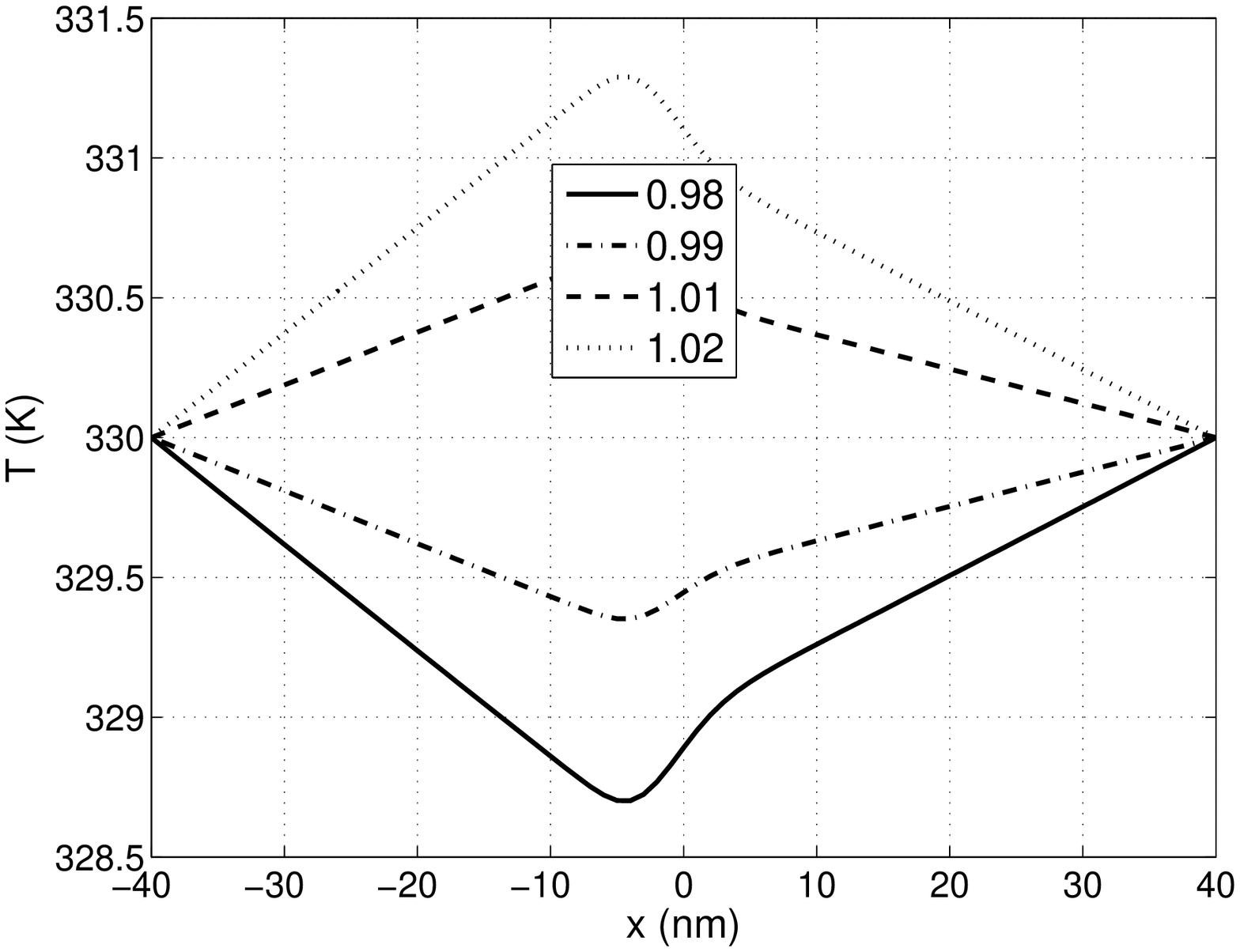}\label{Fig_T_pg} } %
\subfigure[Chemical potential ] %
{\includegraphics[scale=\profilescale]{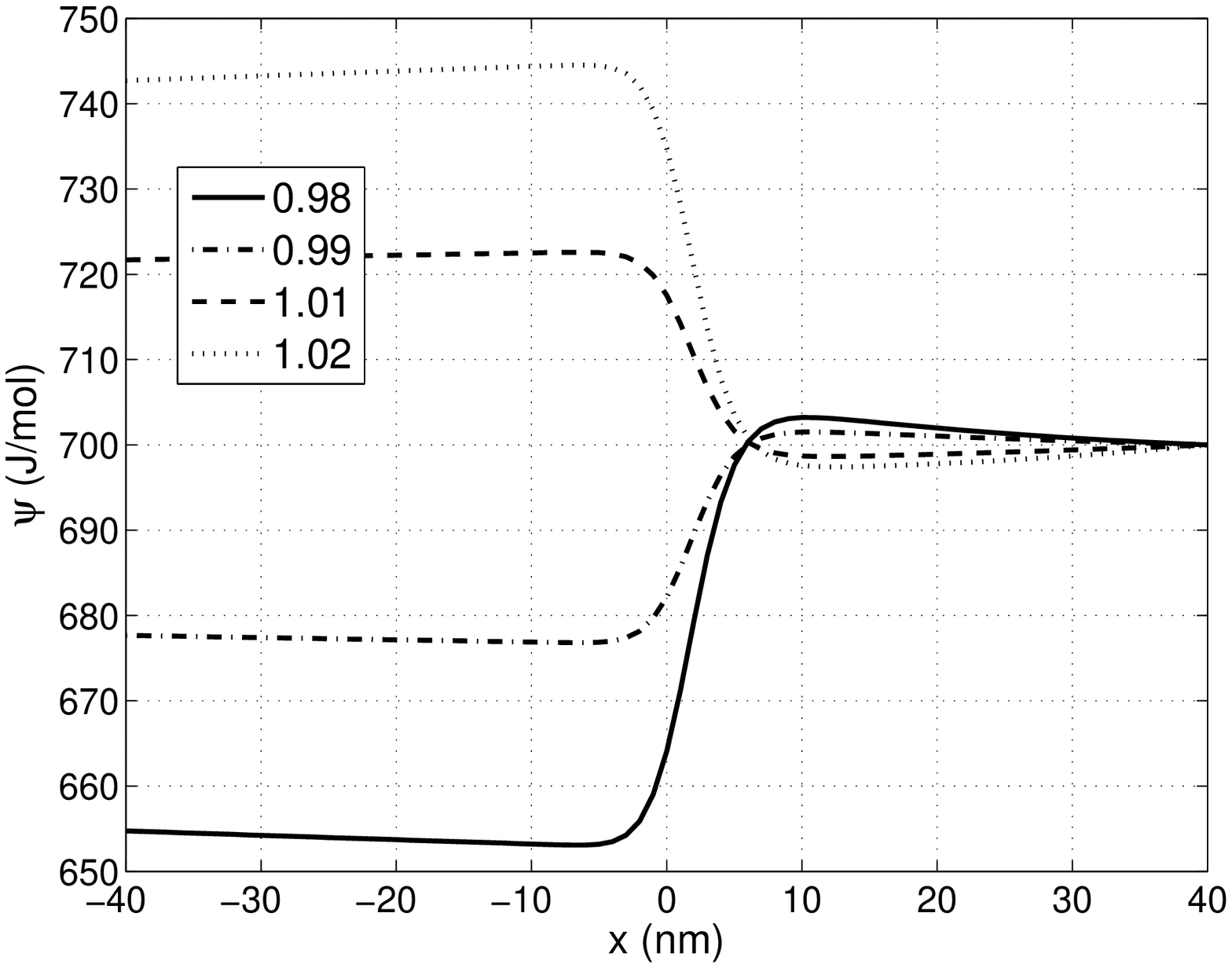} \label{Fig_psi_pg} } %
\caption{Temperature and chemical potential profiles at various $p^{g}$}\label{Fig_pg}
\end{figure}

\begin{figure}
\centering
\subfigure[Temperature  ] %
{\includegraphics[scale=\profilescale]{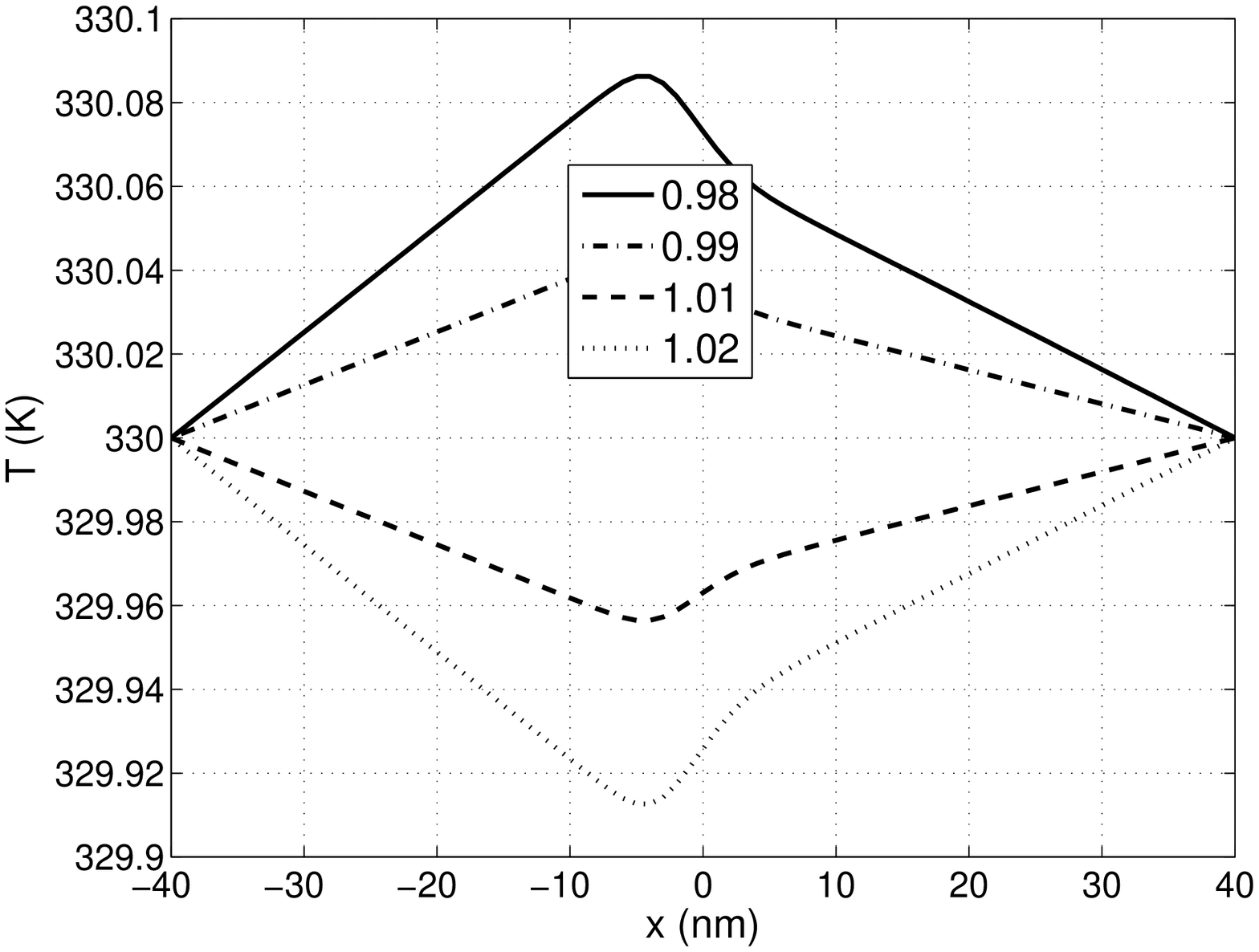}\label{Fig_T_zil} } %
\subfigure[Chemical potential ] %
{\includegraphics[scale=\profilescale]{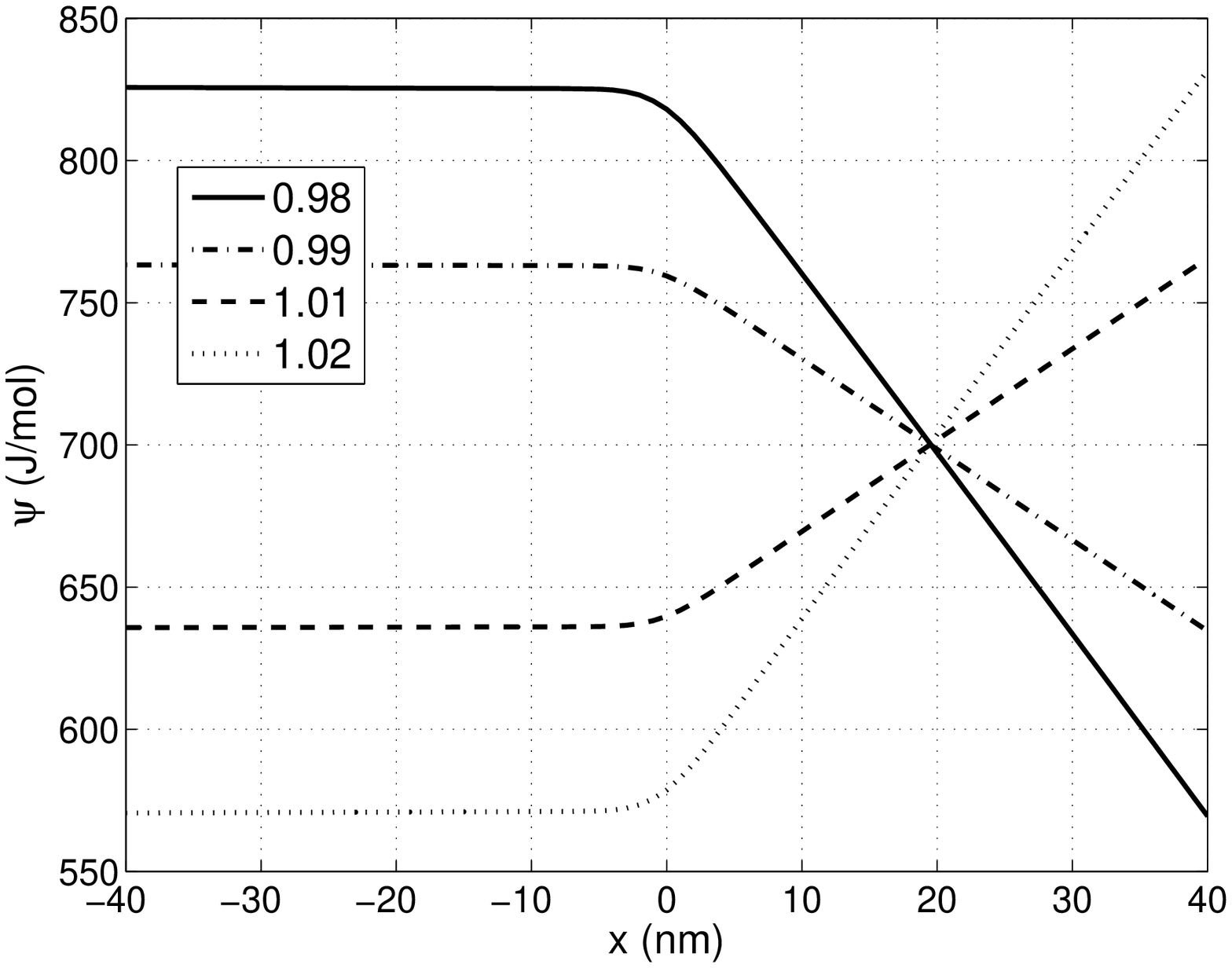} \label{Fig_psi_zil} } %
\caption{Temperature and chemical potential profiles at various $\zeta^{\ell}$}\label{Fig_zil}
\end{figure}

The mixture is then perturbed from equilibrium for the following three cases: 1) setting $T^{\ell}$ equal to 0.98, 0.99, 1.01, 1.02 of $T_{eq}$ and keeping $T^{g}$, $p^{g}$ and $\zeta^{\ell}$
equal to their equilibrium values, see \figr{Fig_Tl}; 2) setting $p^{g}$ equal to 0.98, 0.99, 1.01, 1.02 of $p_{eq}$ and keeping $T^{g}$, $T^{\ell}$ and $\zeta^{\ell}$ equal to their equilibrium
values, see \figr{Fig_pg}; 3) setting $\zeta^{\ell}$ equal to 0.98, 0.99, 1.01, 1.02 of $\zeta_{eq}^{\ell}$ and keeping $T^{g}$, $T^{\ell}$ and $p^{g}$ equal to their equilibrium values, see
\figr{Fig_zil}.
\section{Local equilibrium of the surface}\label{sec/LocalEquilibrium}

In equilibrium it is possible to describe the surface in terms of Gibbs excess quantities \cite{Gibbs/ScientificPapers}. One can treat a system of coexisting liquid and vapor as a three-phase
system: liquid and vapor bulk phases and the surface phase. The surface has thermodynamic properties. The temperature and chemical potentials have the same equilibrium value as in the rest of
the system. Furthermore the thermodynamic state of the surface is given by excess concentrations and thermodynamic potentials. Following Gibbs we have for the surface
\begin{equation}\label{eq/LocalEquilibrium/01}
\begin{array}{rl}
h_{eq}^{s} &= \mu_{1,eq}\,c_{1,eq}^{s}  + \mu_{2,eq}\,c_{2,eq}^{s} + T_{eq}\,s^{s}_{eq} \\
u_{eq}^{s} &= \mu_{1,eq}\,c_{1,eq}^{s}  + \mu_{2,eq}\,c_{2,eq}^{s} + \gamma^{s}_{eq} + T_{eq}\,s^{s}_{eq} \\
f_{eq}^{s} &= \mu_{1,eq}\,c_{1,eq}^{s}  + \mu_{2,eq}\,c_{2,eq}^{s} + \gamma^{s}_{eq} \\
g_{eq}^{s} &= \mu_{1,eq}\,c_{1,eq}^{s}  + \mu_{2,eq}\,c_{2,eq}^{s} \\
\end{array}
\end{equation}
The superscript $s$ indicates here the surface and equal to the excess of corresponding quantities. In these relations the temperature and the chemical potentials, which are the same everywhere,
are independent of the choice of the dividing surface. The excesses depend on the choice of the dividing surface, in a such way that the above relations are true for any choice of the dividing
surface.

It is our aim in this paper to show that the surface in a non-equilibrium liquid-vapor system can also be described as a separate thermodynamic phase using the Gibbs excess quantities. We will
call this property the local equilibrium of the surface. The property of local equilibrium for the surface implies that it is possible to define all thermodynamic properties of a surface such
that they have their equilibrium coexistence values for any choice of the dividing surface given the temperature of the surface $T^{s}$ and the chemical potential difference $\psi^{s}$. For this
purpose we will define the excess densities and develop a method to obtain $T^{s}$ and $\psi^{s}$ independent of the choice of the dividing surface below. The analysis will be done using the
numerical solution of the system in stationary non-equilibrium states.

\subsection{Defining the excess densities.}\label{sec/LocalEquilibrium/Excess}

The definition of the excesses consists of 3 steps: determining the phase boundaries, defining the specified dividing surface and, in particular, defining the excesses.

To determine the phase boundaries we will use the order parameter $q$. We introduce a small parameter $\beta$ and define the $\beta$-dependent boundary, $\xgsb$ between the vapor and the surface
by
\begin{equation}\label{eq/LocalEquilibrium/Excess/01}
\left\vert \frac{q(\xgsb) - q_{eos}(p_{\perp}(\xgsb),\, \xi(\xgsb),\, T(\xgsb))}{q(\xgsb)}\right\vert \equiv \beta
\end{equation}
where $q_{eos}(p_{\perp},\, \xi,\, T)$ is the equation of state's value (no gradient contributions) of $q$ for pressure $p_{\perp}$, mass fraction $\xi$ and temperature $T$. The
$\beta$-dependent boundary, $\xlsb$, between the surface and the liquid is defined in the same way.

The numerical procedure calculates profiles only at specified grid points, which we provide to the procedure. That means that $\xgsb$ and $\xlsb$ can only be situated at points of the grid. We
choose their position to be the last bulk point of the grid where the left hand side of \eqr{eq/LocalEquilibrium/Excess/01} does not exceed the right hand side. In our calculations we will
choose $\beta=10^{-3}$ and use a grid of 81 points.

We shall also choose bulk boundaries near the box boundary where, because of the finite size of the box, the behavior of the profiles might be uncharacteristic. To avoid this effect, we do not
consider the first 5 points of each phase close to these boundaries when we calculate the properties in these phases. The 6th point we call $\xg$ and the 76th point $\xl$.

The bulk gas therefore ranges from $\xg$ to $\xgsb$ and the bulk liquid ranges from $\xlsb$ to $\xl$. The surface therefore ranges from $\xgsb$ to $\xlsb$. In order to define excess quantities
properly we always choose conditions such, that the bulk widths, $\xgsb-\xg$ and $\xl-\xlsb$ are larger then the surface width $\xlsb-\xgsb$.

In order to determine excess densities we need to extrapolate bulk profiles into the interfacial region. In equilibrium extrapolated bulk profiles are constants which are equal to the coexisting
values of the corresponding quantities. Non-equilibrium bulk profiles are not constant. We fit the bulk profile with a polynomial of order $n_{b}=2$ and use this polynomial to extrapolate
non-equilibrium bulk profiles into the interfacial region. This is done with the help of Matlab functions \texttt{polyfit} and \texttt{polyval}. It is important to realize that the extrapolation
of the bulk profiles introduces a certain error depending on the choice of $\beta$ and $n_{b}$ in particular for non-equilibrium systems.

The distances between commonly used dividing surfaces, such as for instance the equimolar surface and the surface of tension, are very small. Thus, if there occurs an error in the determination
of a dividing surface using a course grid, that would lead to inaccurate results. We therefore divide each interval of the course grid between $\xgsb$ and $\xlsb$ in $10^{4}$ subintervals. This
surface grid is used for all operations related to the surface. Within the interfacial region we interpolate all the profiles (which were obtained by extrapolation from the bulk to the surface
region using the course grid) from the course grid to the surface grid using a polynomial of order $n_{s}=3$ with the help of Matlab functions \texttt{polyfit} and \texttt{polyval}.

We can now define the excess $\widehat{\phi}$ of any density $\phi(x)$ as a function of a dividing surface $\xs$ as
\begin{equation}\label{eq/LocalEquilibrium/Excess/02}
\widehat{\phi}(\xs) = \int_{\displaystyle \xgsb}^{\displaystyle \xlsb}{dx\,[\phi(x) - \phi^{g}(x)\,\Theta(\xs-x) -  \phi^{\ell}(x)\,\Theta(x-\xs)]}
\end{equation}
where $\phi ^{g}$ and $\phi ^{\ell }$ are the extrapolated gas and liquid profiles and $\Theta(t)$ is the Heaviside function. The density $\phi$ is per unit of volume and $\widehat{\phi}$ is per
unit of surface area. In our calculations integration is performed using the trapezoidal method by Matlab function \texttt{trapz}.

We can now define different dividing surfaces. The equimolar dividing surface $x^{c}$ is defined by the equation $\widehat{c}(x^{c})=0$. Analogously we define equimolar surfaces with respect to
component 1 and 2: $\widehat{c}_{1}(x^{c_{1}})=0$ and $\widehat{c}_{2}(x^{c_{2}})=0$, and the equidensity surface $x^{\rho}$. The surface of tension $x^{\gamma}$ is defined from the equation
$x^{\gamma}\,\widehat{p}_{\parallel}(x^{\gamma})-\widehat{x\,p_{\parallel}}(x^{\gamma }) = 0$. All the densities are given as arrays on a coordinate grid, but not as continuous functions. Thus,
in order to find the solution of an equation $\varphi(x)=0$ we calculate the values $\varphi_{i} = \varphi(x_{i})$ for each point $x_{i}$ within the surface region and find the minimum of it's
absolute value, $\min_{i}(|\phi_{i}|)$. Because of the discrete nature of the argument this value may not be equal to zero, but it will be the closest to zero among all other coordinate points.
So we will call this point the root of the equation $\varphi (x)=0$. We use the fine surface grid in this procedure.

If follows from \eqr{eq/LocalEquilibrium/Excess/02} that
\begin{equation}\label{eq/LocalEquilibrium/Excess/03}
\frac{d\,\widehat{\phi}(x_{s})}{d\,x_{s}} = \phi^{\ell}(x_{s}) - \phi^{g}(x_{s})
\end{equation}
which we will use later.

\subsection{Defining the temperature and chemical potential difference}\label{sec/LocalEquilibrium/Temperature}

An equilibrium two-phase two-component mixture has two free parameters, for instance the temperature $T$ and the chemical potential difference $\psi = \mu_{1}-\mu_{2}$. \textit{Local equilibrium
of a surface} implies, that in non-equilibrium it should be possible to define the temperature $T^{s}$ and the chemical potential difference $\psi^{s}$ of the surface. As found in ref.
\cite{bedeaux/vdW/II} for the surface temperature in the one-component system, both $T^{s}$ and $\psi ^{s}$ should be independent of the choice of the dividing surface.

The equilibrium temperature and chemical potential difference determine all other equilibrium properties of the surface. Thus, there is a bijection from $T_{eq}$ and $\psi_{eq}$ to any other set
of independent excess variables $X_{1,eq}$ and $X_{2,eq}$, so that one can use them equally well in order to characterize a surface. In non-equilibrium the temperature and chemical potential
difference vary through the interfacial region, but as $X_{1,ne}$ and $X_{2,ne}$ are excesses, they characterize the whole surface. If a non-equilibrium surface is in local equilibrium, there
should exist the same bijection. This implies that given two independent non-equilibrium excesses $X_{1,ne}$ and $X_{2,ne}$ one can determine the temperature $T^{s}$ and the chemical potential
$\psi^{s}$ of the whole surface. Thus one can calculate equilibrium tables of $X_{1,eq}$($T_{eq}$, $\psi_{eq}$) and $X_{2,eq}$($T_{eq}$, $\psi_{eq}$) for different values of $T_{eq}$ and
$\psi_{eq}$ and then determine temperature and chemical potential of a surface as $T^{s} = T_{eq}$($X_{1,ne}$, $X_{2,ne}$) and $\psi^{s}=\psi_{eq}$($X_{1,ne}$, $X_{2,ne}$).

As we want the temperature and chemical potential difference to be independent of the position of the dividing surface, we shall use excesses which are also independent of the position of a
dividing surface in equilibrium for $X_{1}$ and $X_{2}$. For two component mixture these independent variables are the surface tension $\gamma$ and the relative adsorption $\Gamma_{12}$. If the
number of components is more then 2, additional relative adsorptions should be used.

These quantities are well defined for equilibrium, but not for non-equilibrium. So we will define them first. In equilibrium the surface tension is defined as minus the excess of the parallel
pressure $\gamma_{eq} = -\widehat{p}_{\parallel}$. Alternatively one often uses the integral of $p_{\perp}-p_{\parallel }(x)\equiv \gamma_{xx}(x)$ across the interface:
$\gamma_{eq}=\int{dx\,\gamma_{xx}(x)}$. Both definitions are equivalent in equilibrium since $p_{\perp}$ is constant through the interface and $\gamma_{xx}(x)$ is identically zero in the bulk
phases. In non-equilibrium $\gamma_{xx}(x)$ may differ from zero in the bulk regions, however, and this makes the second definition inappropriate. We will therefore define the non-equilibrium
surface tension using the standard definition
\begin{equation}\label{eq/LocalEquilibrium/Temperature/03}
\gamma(\xs) = - \widehat{p}_{\parallel}(\xs)
\end{equation}
This quantity differs from $\widehat{\gamma}_{xx}$ by the term equal to $\widehat{p}_{\perp}$, which is usually small compared to $\widehat{p}_{\parallel}$.

The relative adsorption is defined as $\Gamma_{12, eq} = \widehat{c}_{1, eq}-\widehat{c}_{2, eq}\,(c^{\ell}_{1, eq}-c^{g}_{1, eq})/(c^{\ell}_{2, eq}-c^{g}_{2, eq})$ in equilibrium
\cite{DefayPrigogine/sta}, where $c^{\ell}_{i, eq}$ and $c^{g}_{i, eq}$ are coexistence concentrations of the corresponding components. Since these quantities are not constant in
non-equilibrium, we cannot use this definition directly. One can however see from \eqr{eq/LocalEquilibrium/Excess/03}, that both, in equilibrium and non-equilibrium $\widehat{c}_{i}\,'(\xs) =
c^{\ell}_{i}(\xs)-c^{g}_{i}(\xs)$, where the prime indicates a spatial derivative. Since in equilibrium $c^{\ell}_{i}(\xs)-c^{g}_{i}(\xs) = c^{\ell}_{i, eq}-c^{g}_{i, eq}$ we can use the
following definition
\begin{equation}\label{eq/LocalEquilibrium/Temperature/04}
\Gamma_{12}(\xs) = \widehat{c}_{1}(\xs) - \widehat{c}_{2}(\xs)\,{\frac{c^{\ell}_{1}(\xs)-c^{g}_{1}(\xs)}{c^{\ell}_{2}(\xs)-c^{g}_{2}(\xs)}}
\end{equation}
both in equilibrium and non-equilibrium.

If the system is in local equilibrium we may write:
\begin{equation}\label{eq/LocalEquilibrium/Temperature/05}
\begin{array}{rl}
\gamma(\xs) &= \gamma_{eq}(T^{s}, \psi^{s})\\
\Gamma_{12}(\xs) &= \widehat{c}_{1}(\xs) - \widehat{c}_{2}(\xs)\,\displaystyle{\frac{c^{\ell}_{1, eq}(T^{s},\psi^{s})-c^{g}_{1, eq}(T^{s},\psi^{s})}{c^{\ell}_{2, eq}(T^{s},\psi^{s})-c^{g}_{2,
eq}(T^{s},\psi^{s})}}
\end{array}
\end{equation}
Substituting the expressions for $\gamma(x_{s})$ and $\Gamma_{12}(x_{s})$ from \eqr{eq/LocalEquilibrium/Temperature/03} and \eqr{eq/LocalEquilibrium/Temperature/04} into
\eqr{eq/LocalEquilibrium/Temperature/05} we obtain the following relations
\begin{equation}\label{eq/LocalEquilibrium/Temperature/06}
\begin{array}{rl}
\widehat{p}_{\parallel}(\xs) &= \widehat{p}_{\parallel, eq}(T^{s}, \psi^{s})\\ \\
\displaystyle{\frac{c^{\ell}_{1}(\xs)-c^{g}_{1}(\xs)}{c^{\ell}_{2}(\xs)-c^{g}_{2}(\xs)}} &= \displaystyle{\frac{c^{\ell}_{1, eq}(T^{s},\psi^{s})-c^{g}_{1, eq}(T^{s},\psi^{s})}{c^{\ell}_{2,
eq}(T^{s},\psi^{s})-c^{g}_{2, eq}(T^{s},\psi^{s})}}
\end{array}
\end{equation}
This gives the bijection equations to determine $T^{s}$ and $\psi^{s}$ from the actual non-equilibrium variables. As the left hand sides in \eqr{eq/LocalEquilibrium/Temperature/06} are in good
approximation independent of the position of the dividing surface, $T^{s}$ and $\psi^{s}$ are similarly independent on this position.

\subsection{Defining local equilibrium}\label{sec/LocalEquilibrium/Define}

The other quantities required for the Gibbs description of the non-equilibrium surfaces we define in the following way. The surface chemical potentials are the equilibrium coexistence values
determined via the procedure discussed in \ssecr{sec/Procedure/Equilibrium}
\begin{equation}\label{eq/LocalEquilibrium/Define/01}
\begin{array}{rl}
\mu_{1}^{s} &\equiv \mu_{1,\,eq}(T^{s}, \psi^{s})\\
\mu_{2}^{s} &\equiv \mu_{2,\,eq}(T^{s}, \psi^{s})\\
\end{array}
\end{equation}
We define the surface extensive properties as\footnote{Note, that for some quantities this definition differs from the one, used in \cite{bedeaux/vdW/II}. We will come back to this point later.}
\begin{equation}\label{eq/LocalEquilibrium/Define/02}
\phi^{s}(\xs) \equiv \widehat{\phi}(\xs)
\end{equation}

The local equilibrium of a surface should be established for any choice of a dividing surface. The results of the calculations for any particular choice of a dividing surface may not be
representative since they may be different differ for another choice of a dividing surface. Thus, the property of local equilibrium should be established for all dividing surfaces together.

\begin{figure}[hbt!]
\centering
\includegraphics[scale=\figscale]{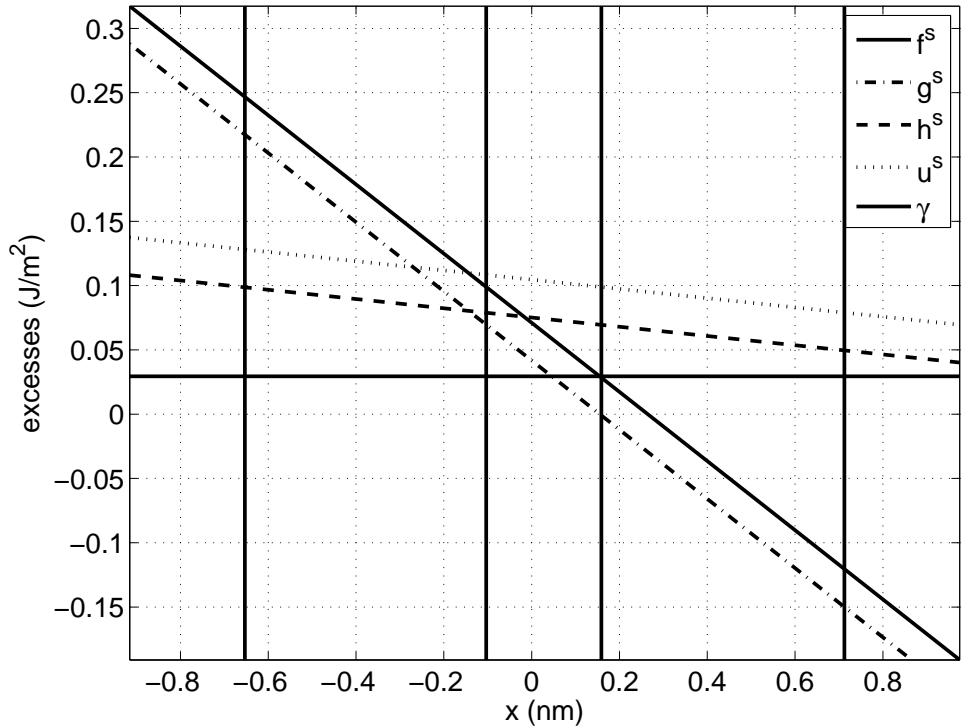}
\caption{Equilibrium excesses at $T=330$ K and $\psi=700$ J/mol. The vertical lines indicate the $x^{c_{2}}$, $x^{\gamma}$, $x^{c}$, $x^{c_{1}}$ dividing surfaces from left to
right.}\label{Eq_Excesses}
\end{figure}
\begin{figure}[hbt!]
\centering
\includegraphics[scale=\figscale]{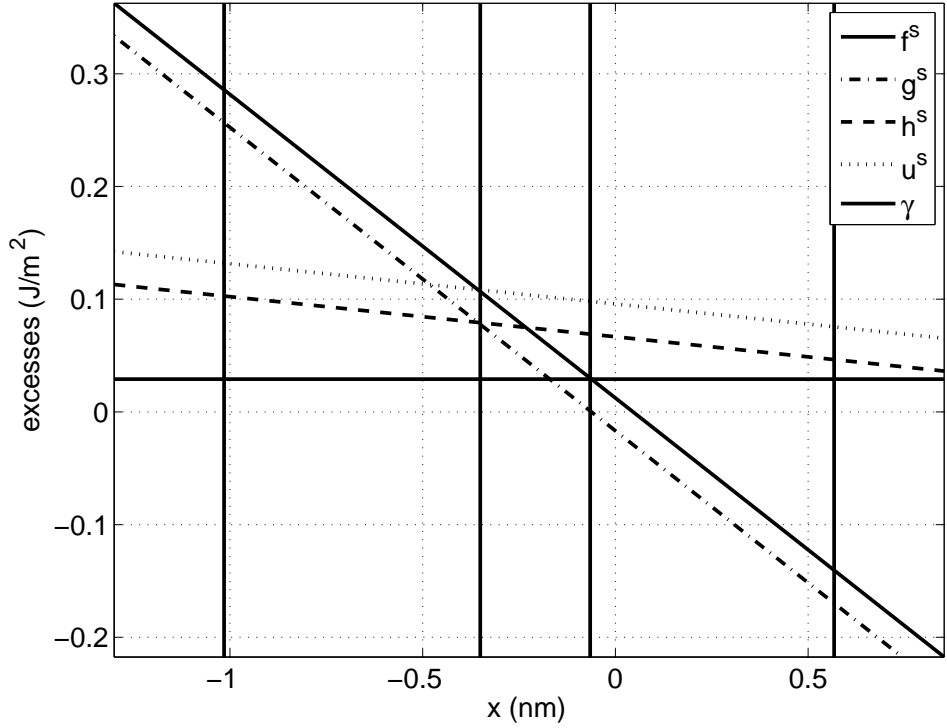}
\caption{Non-equilibrium excesses for the case of perturbing $T^{\ell} = 1.02\,T_{eq}$. The vertical lines indicate the $x^{c_{2}}$, $x^{\gamma}$, $x^{c}$, $x^{c_{1}}$ dividing surfaces from
left to right.}\label{NonEqTl102_Excesses}
\end{figure}
\begin{figure}[hbt!]
\centering
\includegraphics[scale=\figscale]{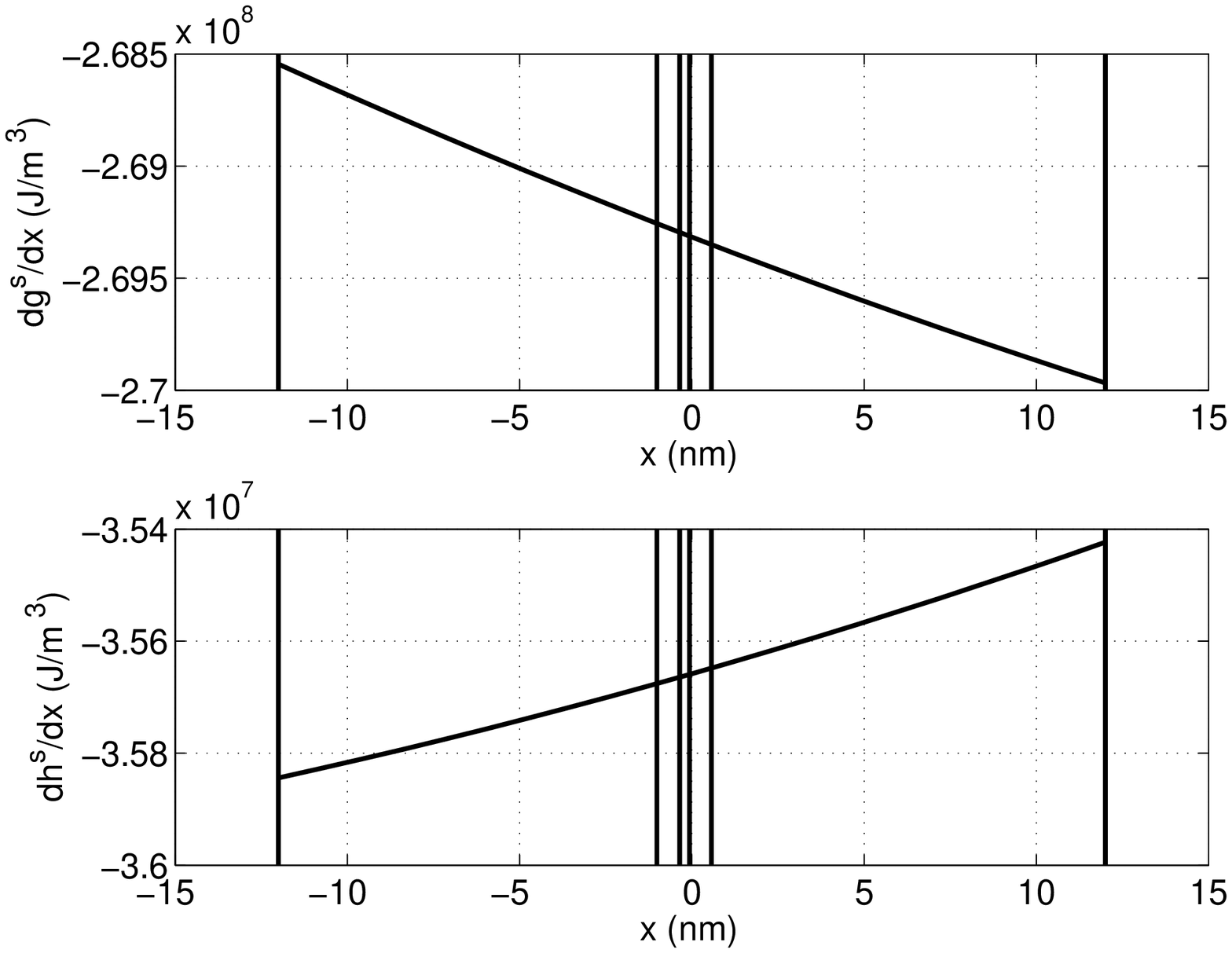}
\caption{Slopes of non-equilibrium excesses for the case of perturbing $T^{\ell} = 1.02\,T_{eq}$. The vertical lines indicate the surface boundaries and $x^{c_{2}}$, $x^{\gamma}$, $x^{c}$,
$x^{c_{1}}$ dividing surfaces.}\label{NonEqTl102_Diff_Excesses}
\end{figure}

Consider the profile of an excess $\widehat{\phi}(\xs)$ as a function of position of a dividing surface $\xs$. It follows from \eqr{eq/LocalEquilibrium/Excess/03} that the slope of the excess
profile is equal to the difference between the extrapolated values of a profile of thermodynamic quantity $\phi$. In equilibrium these values are constant and equal to the coexistence values.
Thus, equilibrium excess profiles are straight lines, as one can see on \figr{Eq_Excesses}. Non-equilibrium profiles in the bulk phases are not constant. We construct the extrapolated profiles
using $n_{b}$th order polynomials with $n_{b}=2$. Resulting non-equilibrium excesses are therefore polynomials of the order $n_{b}+1 = 3$, according to \eqr{eq/LocalEquilibrium/Excess/03}. These
profiles, for the most extreme case of non-equilibrium perturbation $T^{\ell} = 1.02\,T_{eq}$, are shown in \figr{NonEqTl102_Excesses}. Even though these profiles are polynomials of the 3rd
order they are very close to straight lines. As one can see from \figr{NonEqTl102_Diff_Excesses} the variation in the slope is about 1\% through the whole surface. It indicates that this
non-equilibrium "state" is very close to an equilibrium one.

We therefore develop the procedure to relate the non-equilibrium state to an equilibrium one by comparing thermodynamic quantities in equilibrium and in non-equilibrium. The comparison performed
in one particular point of the surface may not be sufficient because it may suffer from artefacts peculiar to this particular surface. Moreover, any comparison performed in a particular point
does not speak for the whole. We therefore compare the non-equilibrium surface with an equilibrium one for all dividing surfaces together. We will use the least square sum method for this.

Consider a non-equilibrium thermodynamic excess $t^{s}(\xs)$ and a quantity $r^{s}(\xs; T, \psi)$ which is a combination of excesses and may depend on $(T, \psi)$ as parameters. We introduce the
following measures of the difference of $t^{s}$ and $r^{s}$
\begin{equation}\label{eq/LocalEquilibrium/Define/03}
\delta_{t^{\ssy s},r^{\ssy s}}(\xs; T, \psi) \equiv |t^{s}(\xs_{i}) - r^{s}(\xs_{i}; T, \psi)|
\end{equation}
and
\begin{equation}\label{eq/LocalEquilibrium/Define/04}
\begin{array}{rl}
S_{t^{\ssy s},r^{\ssy s}}(T, \psi) &\equiv \displaystyle \sum_{i\,\in\, surface}{\left[t^{s}(\xs_{i}) - r^{s}(\xs_{i}; T, \psi)\right]^{2}} \\
\\
\sigma_{t^{\ssy s},r^{\ssy s}}(T, \psi) &\equiv \displaystyle {1 \over N}\,\sqrt{S_{t^{\ssy s},r^{\ssy s}}(T, \psi)}
\end{array}
\end{equation}
where $N$ is the number of surface points.

We say that $t^{s}$ and $r^{s}$ are \textit{the same} in the surface if the value of $\sigma_{t^{\ssy s},r^{\ssy s}}(T, \psi)$ is negligible compared to the typical value of either
$|t^{s}(\xs)|$ or $|r^{s}(\xs; T, \psi)|$. If two quantities $t^{s}$ and $r^{s}$ are the same in the above sense, we say that the non-equilibrium state of the surface is characterized by surface
temperature $T^{s}_{\xxs}$ and chemical potential difference $\psi^{s}_{\xxs}$ if
\begin{equation}\label{eq/LocalEquilibrium/Define/04a}
S_{t^{\ssy s},r^{\ssy s}}(T^{s}_{\xxs}, \psi^{s}_{\xxs}) = \min_{T, \psi} S_{t^{\ssy s},r^{\ssy s}}(T, \psi)
\end{equation}
Here superscript $s$ indicates that we speak about surface quantities only (as everywhere in this paper) and subscript $\xxs$ indicates that $T^{s}_{\xxs}$ and $\psi^{s}_{\xxs}$ are the
parameters for all dividing surfaces together (in contrast to the values $T^{s}(\xs)$ and $\psi^{s}(\xs)$ determined from \eqr{eq/LocalEquilibrium/Temperature/06} for each particular dividing
surface $\xs$).

These definitions are easy to illustrate in equilibrium. For instance, for
$h^{s}_{eq,\,gibbs}(\xs; T, \psi) \equiv \mu_{1,eq}(T, \psi)\,c_{1,eq}^{s}  + \mu_{2,eq}(T, \psi)\,c_{2,eq}^{s} + T\,s^{s}_{eq}$ %
it follows from \eqr{eq/LocalEquilibrium/01}, that
$h^{s}_{eq}(\xs; T_{eq}, \psi_{eq}) = h^{s}_{eq,\,gibbs}(\xs; T_{eq}, \psi_{eq})$%
. Furthermore
$g^{s}_{eq}(\xs; T_{eq}, \psi_{eq}) \neq h^{s}_{eq,\,gibbs}(\xs; T_{eq}, \psi_{eq})$%
. Thus
$\delta_{h^{\ssy s}_{\vphantom{\ssy gibbs}\ssy eq},\,h^{\ssy s}_{\ssy eq,\,gibbs}}(\xs; T_{eq}, \psi_{eq}) = 0$ %
and
$\delta_{g^{\ssy s}_{\vphantom{\ssy gibbs}\ssy eq},\,h^{\ssy s}_{\ssy eq,\,gibbs}}(\xs; T_{eq}, \psi_{eq}) \neq 0$%
. It is also true that
$S_{h^{\ssy s}_{\vphantom{\ssy gibbs}\ssy eq},\,h^{\ssy s}_{\ssy eq,\,gibbs}}(T_{eq}, \psi_{eq}) = \min S_{h^{\ssy s}_{\vphantom{\ssy gibbs}\ssy eq},\,h^{\ssy s}_{\ssy eq,\,gibbs}}(T, \psi)$ %
and
$\sigma_{h^{\ssy s}_{\vphantom{\ssy gibbs}\ssy eq},\,h^{\ssy s}_{\ssy eq,\,gibbs}}(T, \psi) \ll \sigma_{g^{\ssy s}_{\vphantom{\ssy gibbs}\ssy eq},\,h^{\ssy s}_{\ssy eq,\,gibbs}}(T, \psi)$%
. According to the above definitions i) $h^{s}_{eq,\,gibbs}$ and $h^{s}_{eq}$ are the same quantities, but $h^{s}_{eq,\,gibbs}$ and $g^{s}_{eq}$ are not the same; ii) the equilibrium state is
characterized by $(T_{eq}, \psi_{eq})$; as it should be. While this analysis is trivial in equilibrium, it is not trivial in non-equilibrium.

Note, that while in equilibrium the conditions $\delta_{t,r}(\xs; T_{eq}, \psi_{eq}) = 0$ and $S_{t,r}(T_{eq}, \psi_{eq}) = \min S_{t,t}(T, \psi)$ are equivalent, in general it does not follow
in non-equilibrium from \eqr{eq/LocalEquilibrium/Define/04a} that
\begin{equation}\label{eq/LocalEquilibrium/Define/03a}
\delta_{t,r}(\xs; T, \psi) = 0 \quad \forall \; \xs
\end{equation}
Thus \eqr{eq/LocalEquilibrium/Define/03a} is not a good measure of the equality of the quantities and states in non-equilibrium.  We may therefore speak about the equality of thermodynamic
quantities as well as about the state $T^{s}$ and $\psi^{s}$ of the surface in non-equilibrium only in the least square sense, as it is given in \eqr{eq/LocalEquilibrium/Define/04a}.

Within establishing the local equilibrium property of a non-equilibrium surface we would like to verify the following properties: i) the existence of the unique temperature $T^{s}$ and chemical
potential difference $\psi^{s}$ of a non-equilibrium surface; ii) the validity of the \eqr{eq/LocalEquilibrium/01} in non-equilibrium at the surface's $T^{s}$ and $\psi^{s}$; iii) the
possibility to determine all the properties of a non-equilibrium surface from equilibrium tables at the surface's $T^{s}$ and $\psi^{s}$. We do this in the following section.

\section{Verification of local equilibrium.}\label{sec/Results/LocalEquilibrium}

We calculate the the equilibrium properties (coexistence data, such as the pressure or bulk densities, as well as various excesses) of the system for the range of temperatures $T = \{325, 326,
\ldots, 340\}$ K and the range of chemical potentials $\psi = \{400, 450, \ldots, 1000\}$ J/mol. The value of a thermodynamic quantity at any point $(T, \psi)$, which is between these is
interpolated using the Matlab procedures \texttt{interp2} and \texttt{griddata}.

%
\subsection{Surface temperature and chemical potential difference}\label{sec/Results/LocalEquilibrium/Temperature}

As was mentioned, in equilibrium both $\gamma $ and $\Gamma_{12}$ are independent of the location of the dividing surface $\xs$. Given the above definitions,
\eqr{eq/LocalEquilibrium/Temperature/03} and \eqr{eq/LocalEquilibrium/Temperature/04}, we can calculate these quantities for non-equilibrium states. Calculations show, that even though $\gamma$
and $\Gamma_{12}$ are not exactly independent on $\xs$ away from equilibrium, the relative deviation is so small (about $0.004 \%$ for $\gamma$ and $4 \%$ for $\Gamma_{12}$ in the worst case),
that one can consider these quantities to be independent of the position of the dividing surface. Thus one may use them in order to find the temperature, $T^{s}$, and the chemical potential
difference, $\psi^{s}$, of the surface in non-equilibrium states, which will be independent of the position of the dividing surface.

Using \eqr{eq/LocalEquilibrium/Define/04} and \eqr{eq/LocalEquilibrium/Temperature/06} together with \eqr{eq/LocalEquilibrium/Excess/03} we construct the following expressions
\begin{equation}\label{eq/Results/LocalEquilibrium/Temperature/01}
\begin{array}{rl}
S_{\gamma}(T,\psi) &= \displaystyle\sum_{x^{\scriptstyle s}}{\left[\widehat{p}_{\parallel}(\xs) - \widehat{p}_{\parallel, eq}(T, \psi)\right]^{2}} \\
\\
S_{\Gamma_{12}}(T,\psi) &= \displaystyle\sum_{x^{\scriptstyle s}}{\left[  {\frac{\widehat{c}^{\;\prime}_{1,eq}(\xs; T,\psi)}{\widehat{c}^{\;\prime}_{1}(\xs)}}
                                                                    -{\frac{\widehat{c}^{\;\prime}_{2,eq}(\xs; T,\psi)}{\widehat{c}^{\;\prime}_{2}(\xs)}} \right]^{2}} \\
\end{array}
\end{equation}
where the prime indicates the derivative with respect to $\xs$.

\begin{figure}[hbt!]
\centering
\includegraphics[scale=\figscale]{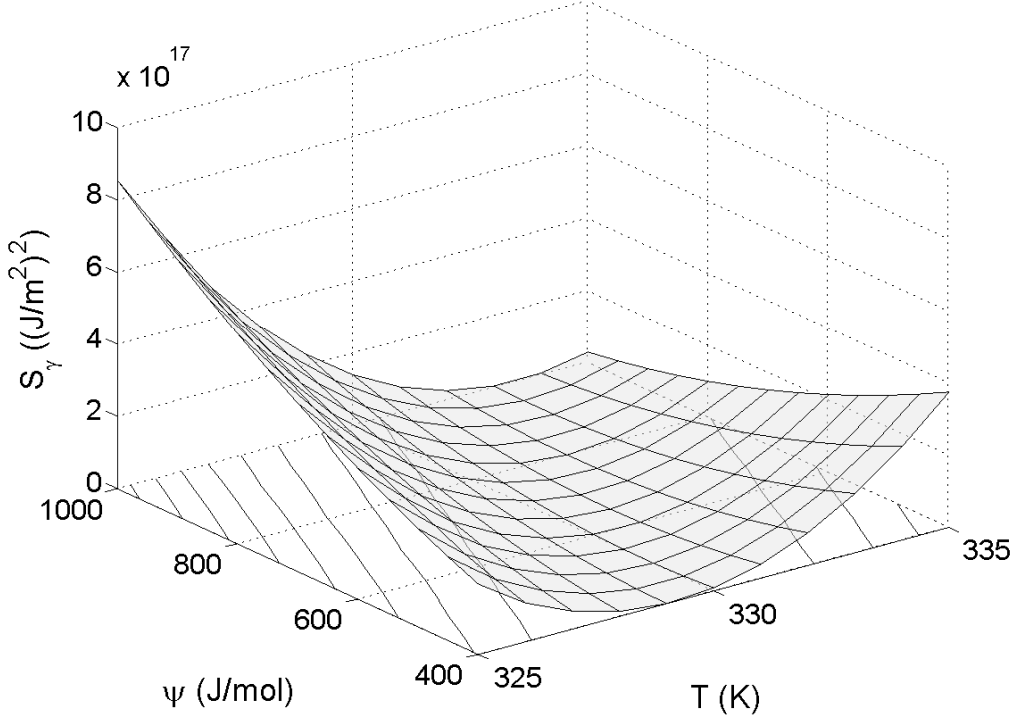}
\caption{The plot of $S_{t}(T, \psi)$ for $t$ being the profile of the surface tension $\gamma$ for the case of perturbing $T^{\ell} = 1.02\,T_{eq}$. The lines in the $T$-$\psi$ plane are lines
of constant value of $S_{t}(T, \psi)$}\label{NonEqTl102_EMC}
\end{figure}

$S_{i}(T,\psi)$ (where $i$ is either $\gamma$ or $\Gamma_{12}$) should reach the minimum at $T^{s}_{\xxs}$ and $\psi^{s}_{\xxs}$. We note however, that neither $S_{i}(T,\psi)$ have a minimum at
a single point $(T^{s},\psi^{s})$. There is a whole generatrix curve of minima $C_{i}(T,\psi)=0$ so the plot of $S_{i}(T,\psi)$ is a valley. One can see it on \figr{NonEqTl102_EMC}. Every point
of the generatrix curve is the minimum point of $S_{i}(T,\psi)$ along the direction "perpendicular" to this generatrix. If $\widehat{p}_{\parallel}(\xs)$ or $\widehat{c}^{\;\prime}_{1}(\xs)$ and
$\widehat{c}^{\;\prime}_{2}(\xs)$ represent the corresponding profiles for some equilibrium state $(T_{eq}, \psi_{eq})$, then $S_{i}(T,\psi) = 0$ and the generatrix is constant. Since these
profiles are non-equilibrium profiles, the generatrix is not exactly constant but very close to it. Thus $C_{i}(T,\psi) = {\partial S_{i}(T,\psi)}/{\partial w}$, where $w$ is a direction in
$T$-$\psi$ plane which is perpendicular to generatrix. In fact, one should be careful speaking about directions, since no metric is defined in the $T$-$\psi$ plane. Thus we cannot introduce
$\nabla_{T\psi}$ so that $C_{i}(T,\psi) = |\nabla_{T\psi}S_{i}(T,\psi)|$. In fact, $w$ can be any direction which does not coincide or does not almost coincide with the direction of generatrix.
In practice we find that we can use $w=T$, while using $w=\psi$ gives less accurate results. Thus we determine the minima curve from the equation
\begin{equation}\label{eq/LocalEquilibrium/Define/05}
\frac{\partial S_{i}(T,\psi)}{\partial T} = 0
\end{equation}

Thus one needs two quantities $S_{\gamma}$ and $S_{\Gamma}$ in order to determine $T^{s}_{\xxs}$ and $\psi^{s}_{\xxs}$ uniquely. The surface temperature and chemical potential difference
$T^{s}_{\xxs}$ and $\psi^{s}_{\xxs}$ are determined from the intersection of two minimum curves of $S_{\gamma}$ and $S_{\Gamma}$
\begin{equation}\label{eq/Results/LocalEquilibrium/Temperature/02}
\begin{array}{rl}
\displaystyle \left. \frac{\partial S_{\gamma}(T,\psi)}{\partial T} \right|_{\,T^{\scriptstyle s}_{\xxs},\,\psi^{s}_{\xxs}} &= 0 \\
\displaystyle \left. \frac{\partial S_{\Gamma_{12}}(T,\psi)}{\partial T}\right|_{\,T^{\scriptstyle s}_{\xxs},\,\psi^{s}_{\xxs}} &= 0 \\
\end{array}
\end{equation}

We calculate the temperatures and the chemical potential differences for different non-equilibrium conditions. They are outlined in \tblsr{tbl/TP/Tl}{tbl/TP/zil}. The first row of each table,
corresponding to $\xxs$, gives $T^{s}$ and $\psi^{s}$ calculated from \eqr{eq/Results/LocalEquilibrium/Temperature/02}. The following rows give, corresponding to different particular dividing
surfaces, gives $T^{s}$ and $\psi^{s}$ calculated from \eqr{eq/LocalEquilibrium/Temperature/06}.
\begin{longtable}{l@{\qquad}l@{\qquad}l@{\qquad\qquad}l@{\qquad}l@{\qquad\qquad}}%
\caption{Surface temperatures (K) and chemical potentials (J/mol) for the case of perturbing $T^{\ell}$} \label{tbl/TP/Tl} \\ %
\hline %
  &  \multicolumn{2}{c}{$T^{\ell} = 1.02\,T_{eq}$}  &  \multicolumn{2}{c}{$T^{\ell} = 0.98\,T_{eq}$}\\ %
\hline %
surface &   $T^{s}$ &   $\psi^{s}$ &   $T^{s}$ &   $\psi^{s}$ \\ %
\hline %
$\xxs$         & 331.831    & 770.53    & 328.129    & 650.92     \\ %
 &  & &  & \\ %
$x^{c}$        & 331.823    & 769.51    & 328.124    & 650.29     \\ %
$x^{\gamma}$   & 331.828    & 770.22    & 328.123    & 650.21     \\ %
$x^{c_{1}}$    & 331.814    & 767.97    & 328.127    & 650.43     \\ %
$x^{c_{2}}$    & 331.838    & 771.86    & 328.121    & 650.1      \\ %
\hline %
\end{longtable} %
\begin{longtable}{l@{\qquad}l@{\qquad}l@{\qquad\qquad}l@{\qquad}l@{\qquad\qquad}}%
\caption{Surface temperatures (K) and chemical potentials (J/mol) for the case of perturbing $p^{g}$} \label{tbl/TP/pg} \\ %
\hline %
  &  \multicolumn{2}{c}{$p^{g} = 1.02\,p_{eq}$}  &  \multicolumn{2}{c}{$p^{g} = 0.98\,p_{eq}$}\\ %
\hline %
surface &   $T^{s}$ &   $\psi^{s}$ &   $T^{s}$ &   $\psi^{s}$ \\ %
\hline %
$\xxs$         & 330.796    & 683.87    & 329.059    & 696.52     \\ %
 &  & &  & \\ %
$x^{c}$        & 330.8      & 684.68    & 329.063    & 697.22     \\ %
$x^{\gamma}$   & 330.799    & 684.44    & 329.063    & 697.12     \\ %
$x^{c_{1}}$    & 330.804    & 685.19    & 329.065    & 697.46     \\ %
$x^{c_{2}}$    & 330.795    & 683.93    & 329.061    & 696.87     \\ %
\hline %
\end{longtable} %
\begin{longtable}{l@{\qquad}l@{\qquad}l@{\qquad\qquad}l@{\qquad}l@{\qquad\qquad}}%
\caption{Surface temperatures (K) and chemical potentials (J/mol) for the case of perturbing $\zeta^{\ell}$} \label{tbl/TP/zil} \\ %
\hline %
  &  \multicolumn{2}{c}{$\zeta^{\ell} = 1.02\,\zeta^{\ell}_{eq}$}  &  \multicolumn{2}{c}{$\zeta^{\ell} = 0.98\,\zeta^{\ell}_{eq}$}\\ %
\hline %
surface &   $T^{s}$ &   $\psi^{s}$ &   $T^{s}$ &   $\psi^{s}$ \\ %
\hline %
$\xxs$         & 329.577    & 559.32    & 330.24     & 812.86     \\ %
 &  & &  & \\ %
$x^{c}$        & 329.598    & 562.44    & 330.242    & 813.16     \\ %
$x^{\gamma}$   & 329.584    & 560.5     & 330.253    & 814.67     \\ %
$x^{c_{1}}$    & 329.63     & 566.83    & 330.219    & 809.99     \\ %
$x^{c_{2}}$    & 329.554    & 556.3     & 330.278    & 817.99     \\ %
\hline %
\end{longtable} %

Note, that $T^{s}$ and $\psi^{s}$ may be different from the continuous values in the interfacial region.


%
\subsection{The non-equilibrium Gibbs surface}\label{sec/Results/LocalEquilibrium/Gibbs}

In this section we would like to verify that the surface quantities defined by  \eqr{eq/LocalEquilibrium/Define/02} satisfy \eqr{eq/Results/LocalEquilibrium/Gibbs/01a} with $T^{s}$ and
$\psi^{s}$ determined by \eqr{eq/LocalEquilibrium/Temperature/06} and \eqr{eq/Results/LocalEquilibrium/Temperature/02}.
\begin{equation}\label{eq/Results/LocalEquilibrium/Gibbs/01a}
\phi^{s} = \phi_{gibbs}^{s}(T^{s}, \psi^{s})
\end{equation}
namely, with the definition \eqr{eq/LocalEquilibrium/Define/01},
\begin{equation}\label{eq/Results/LocalEquilibrium/Gibbs/01}
\begin{array}{rl}
h^{s} &= \mu_{1}^{s}\,c_{1}^{s}  + \mu_{2}^{s}\,c_{2}^{s} + T^{s}\,s^{s} \\
u^{s} &= \mu_{1}^{s}\,c_{1}^{s}  + \mu_{2}^{s}\,c_{2}^{s} + \gamma^{s} + T^{s}\,s^{s} \\
f^{s} &= \mu_{1}^{s}\,c_{1}^{s}  + \mu_{2}^{s}\,c_{2}^{s} + \gamma^{s} \\
g^{s} &= \mu_{1}^{s}\,c_{1}^{s}  + \mu_{2}^{s}\,c_{2}^{s} \\
\end{array}
\end{equation}
where the right hand side is $\phi_{gibbs}^{s}(T^{s}, \psi^{s})$ of the corresponding quantity. \eqr{eq/Results/LocalEquilibrium/Gibbs/01} is the non-equilibrium analogon of equilibrium
\eqr{eq/LocalEquilibrium/01}.

In order to analyze the measure of validity of \eqr{eq/Results/LocalEquilibrium/Gibbs/01a} we construct the quantity
\begin{equation}\label{eq/Results/LocalEquilibrium/Gibbs/02}
\begin{array}{rl}
\displaystyle \EuScript{E}_{\phi_{gibbs}}(T, \psi) &= \displaystyle \sum_{i\,\in\, surface}{\left[\frac{\phi^{s}(\xs_{i}) - \phi^{s}_{gibbs}(\xs_{i};\, T, \psi)}{\phi^{s}(\xs_{i})}\right]^{2}} \\
\\
\displaystyle \epsilon_{\phi_{gibbs}}(\xs; T, \psi) &= \displaystyle \left| \frac{\phi^{s}(\xs)-\phi^{s}_{gibbs}(\xs; T, \psi)}{\phi^{s}(\xs)}\right|
\end{array}
\end{equation}
for each thermodynamic potential $h$, $u$, $f$, $g$. $\EuScript{E}_{\phi_{gibbs}}$ gives the relative error of the determination of the surface quantity $\phi^{s}$ using the Gibbs excesses
relations \eqr{eq/Results/LocalEquilibrium/Gibbs/01} for for all dividing surfaces together, while $\epsilon_{\phi_{gibbs}}$ gives this error for particular dividing surface. We build
$\EuScript{E}_{\phi_{gibbs}}(T, \psi)$ for $T = T^{s}_{\xxs}$, $\psi = \psi^{s}_{\xxs}$ determined from \eqr{eq/Results/LocalEquilibrium/Temperature/02} only for the whole surface. We build
$\epsilon_{\phi_{gibbs}}(T, \psi)$ both for $T = T^{s}_{\xxs}$, $\psi = \psi^{s}_{\xxs}$ determined for the whole surface and for $T = T^{s}(\xs)$, $\psi = \psi^{s}(\xs)$ determined from
\eqr{eq/LocalEquilibrium/Temperature/06} for particular dividing surface. The values of the corresponding errors are listed in \tblsr{tbl/Gibbs/Tl}{tbl/Gibbs/zil} in
\ssecr{sec/ExcessErrors/Gibbs}, and are found to be small.

As one can see, there is a variation in the value of the error for the different dividing surfaces. It is caused by two reasons. The first reason for this is slight variation in $T^{s}$ and
$\psi^{s}$ from \tblsr{tbl/TP/Tl}{tbl/TP/zil} for different dividing surfaces. The variation of each excess potential corresponds to the variation of $T^{s}$ and $\psi^{s}$ through these
surfaces. Thus so do the relative errors.

Another factor which influences the value of these errors is the actual value of an excess at a given dividing surface. If it is close to zero, then in the expression for $\epsilon$ the small
value is in denominator and it gives the huge value for the error. Particularly, $g^{s}(x^{c}) \approx 0$ both in equilibrium and in non-equilibrium which makes the row corresponding to $g$ at
$x^{c}$ be uninformative and one should not take into account these data.

\begin{figure}[hbt!]
\centering
\includegraphics[scale=\figscale]{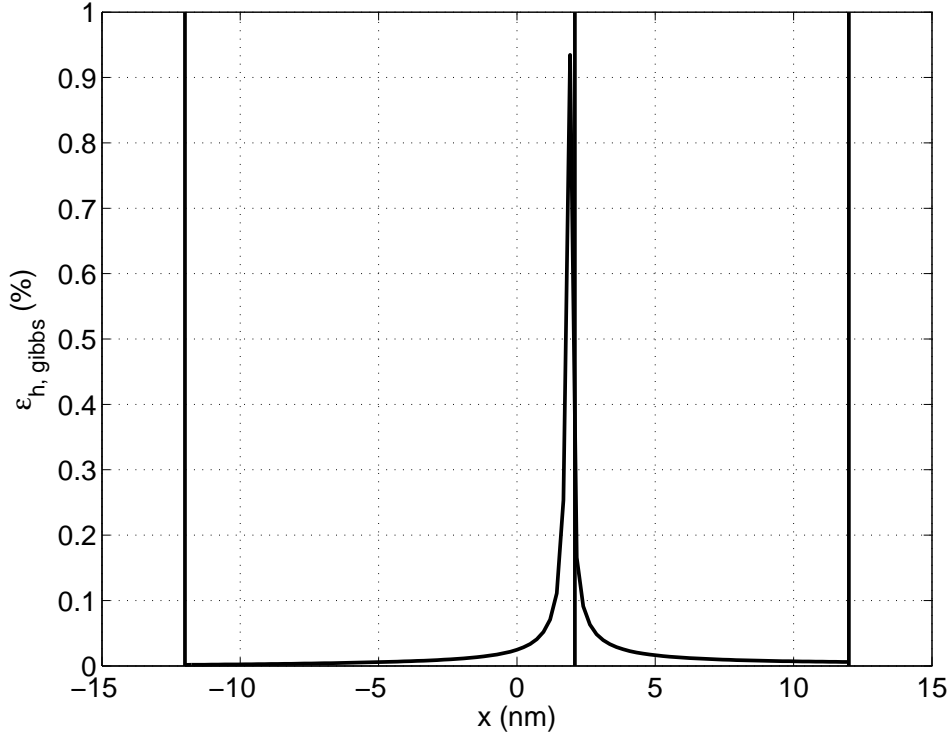}
\caption{The relative error $\epsilon_{h_{gibbs}}(\xs; T^{s}_{\xxs}, \psi^{s}_{\xxs})$ for the case of perturbing $T^{\ell} = 1.02\,T_{eq}$. The vertical lines indicate the surface boundaries
and $x^{h}$ dividing surfaces.}\label{NonEqTl102_Epsilon_Gibbs}
\end{figure}

We emphasize however that the overall error $\EuScript{E}_{\phi_{gibbs}}$ represents the whole surface and thus do not suffer from the fact that some quantity is negligible at some dividing
surface. There are such points for each potential $\phi$, however their contribution to the whole error is negligible itself. So we can see, that if the particular dividing surface is far from
zero point of $\phi$, $\epsilon_{\phi_{gibbs}}$ gives the good measure of the error. While if the particular dividing surface is close to zero point of $\phi$, $\epsilon_{\phi_{gibbs}}$ fails to
measure the error. One can see from \figr{NonEqTl102_Epsilon_Gibbs} that the relative error $\epsilon_{\phi_{gibbs}}$ indeed rises enormously at $x^{\phi}$. Particularly because of this fact the
definition of the excess quantities in \cite{bedeaux/vdW/II} was different from \eqr{eq/LocalEquilibrium/Define/02}.

Another possible test is to compare the absolute error $|\phi^{s}(\xs)-\phi^{s}_{gibbs}(\xs; T, \psi^{s})|$ with the deviation $\sigma_{\phi_{gibbs}}(T, \psi)$ defined in
\eqr{eq/LocalEquilibrium/Define/04}. The calculations show that for the particular dividing surfaces the former quantity does not exceed the lateral, which indicates that all the absolute errors
are actually within the trust region.

\subsection{Equilibrium tables}\label{sec/Results/LocalEquilibrium/Table}

In this subsection we will verify the possibility to determine all the properties of a non-equilibrium surface from equilibrium tables at the surface's $T^{s}$ and $\psi^{s}$. The surface
chemical potentials $\mu_{1}^{s}$ and $\mu_{2}^{s}$ are already defined as their equilibrium values by \eqr{eq/LocalEquilibrium/Define/01}. So in this section we will verify the relation
\begin{equation}\label{eq/Results/LocalEquilibrium/Table/01}
\phi^{s} = \phi^{s}_{eq}(T^{s}, \psi^{s})
\end{equation}

As in \ssecr{sec/Results/LocalEquilibrium/Gibbs} we compare the actual excess of a thermodynamic potential with the corresponding equilibrium value at given temperature and chemical potential of
the surface. Before we do this a note has to be made.

\begin{figure}
\centering
\subfigure[Whole surface] %
{\includegraphics[scale=\profilescale]{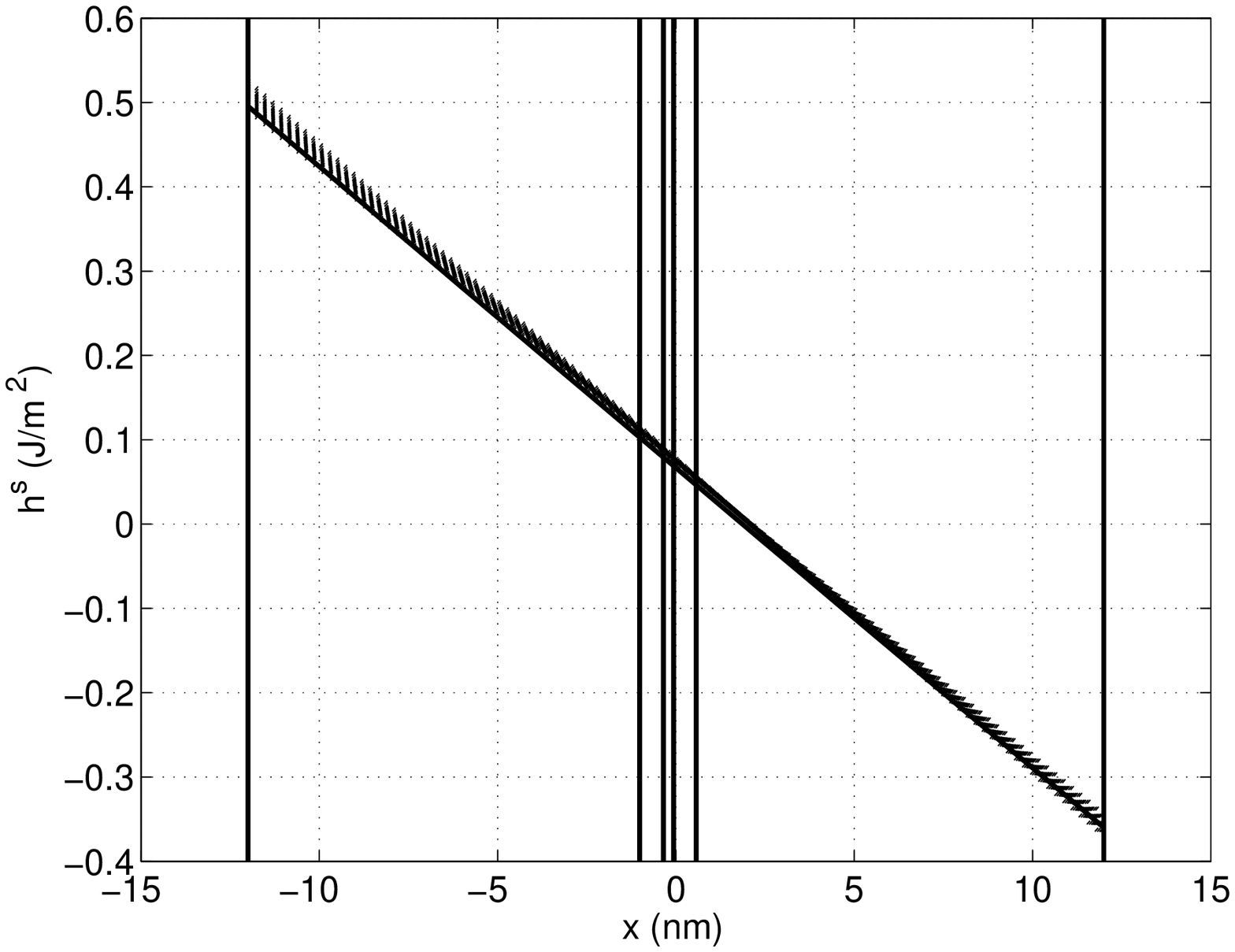}\label{NonEqTl102_Table_Excesses1} } %
\subfigure[Subregion] %
{\includegraphics[scale=\profilescale]{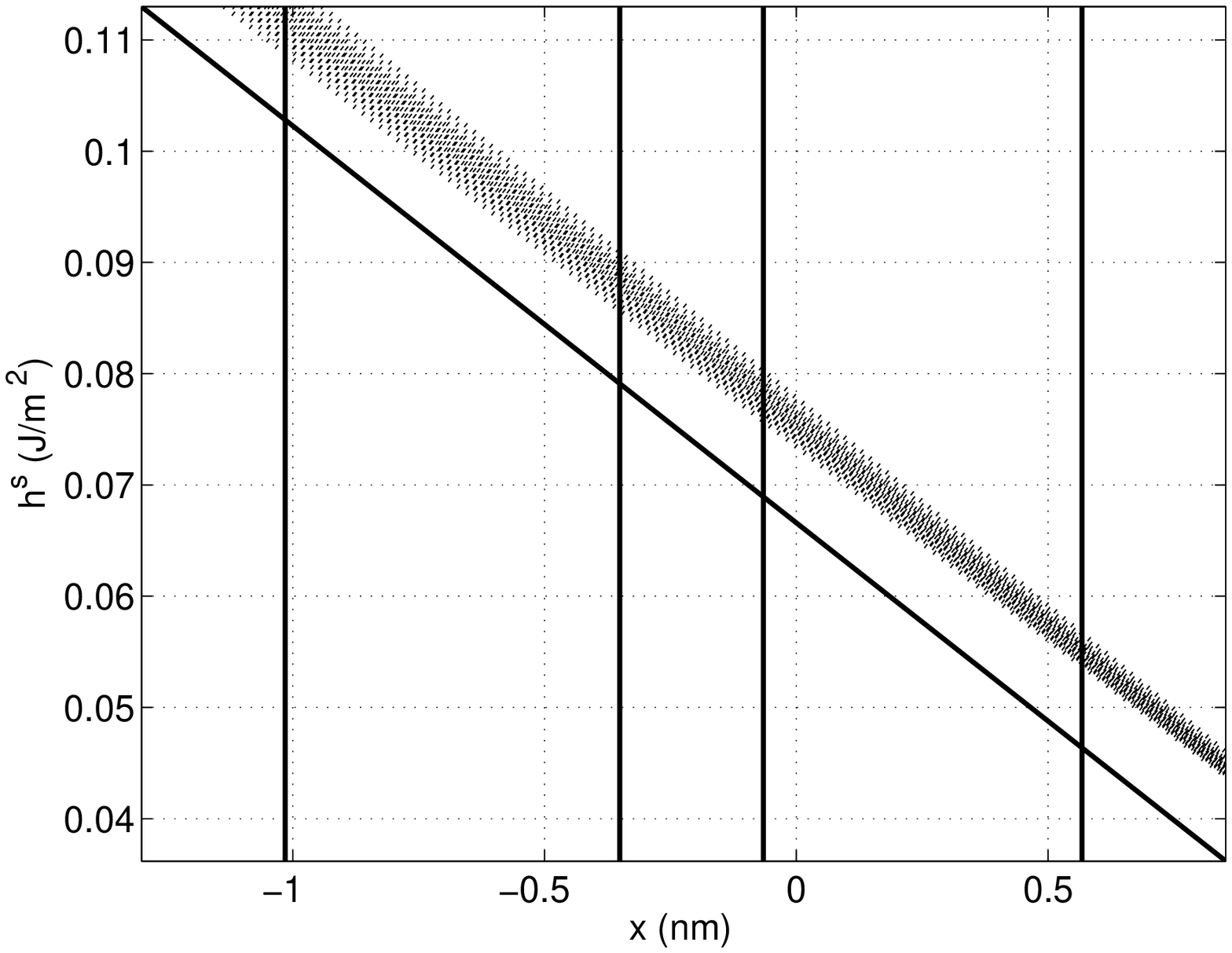} \label{NonEqTl102_Table_Excesses2} } %
\caption{Non-equilibrium profile of $h^{s}$ for the case of perturbing $T^{\ell} = 1.02\,T_{eq}$ (solid line) compared to the profiles of $h^{s}_{eq}$ calculated from the equilibrium tables for
different $T_{eq}$ and $\psi_{eq}$ (dotted lines). The vertical lines indicate the surface boundaries and dividing surfaces for non-equilibrium case.}\label{NonEqTl102_Table_Excesses}
\end{figure}

Under non-equilibrium conditions the profile of a quantity $\phi^{s}$ is shifted with respect to the equilibrium one. One can see this in the example for $\phi=h$ in
\figr{NonEqTl102_Table_Excesses}. The reason for this are fluxes caused by non-equilibrium perturbation. The whole surface is therefore shifted. One can clearly see that comparing the positions
of the particular dividing surfaces on \figr{Eq_Excesses} and \figr{NonEqTl102_Excesses}. So the direct comparison of the profiles should be done not in the observer's frame of reference (OFO,
which is used in all other calculations), but in the \textit{surface's frame of reference} (SFO). The SFO is simply shifted with respect to the OFO, depending on the rate of non-equilibrium
perturbations. Zero of the SFO is chosen at the reference surface, which can be either the equimolar surface, or any other physically sensible surface. If $x^{\ominus}$ is the position of this
surface in OFO and $\phi^{s}_{\mathrm{OFO}}(x^{s}_{\mathrm{OFO}})$ is the profile of $\phi^{s}$ in OFO, then $\phi^{s}_{\mathrm{SFO}}(x^{s}_{\mathrm{SFO}}) \equiv
\phi^{s}_{\mathrm{OFO}}(x^{s}_{\mathrm{OFO}}) = \phi^{s}_{\mathrm{OFO}}(x^{s}_{\mathrm{SFO}}+x^{\ominus})$ is the profile of $\phi^{s}$ in SFO.

We can now determine to which equilibrium state the non-equilibrium one should correspond. Consider the following definitions of $\EuScript{E}_{\phi_{table}}$ and $\epsilon_{\phi_{table}}$ which
have the same meaning as in \eqr{eq/Results/LocalEquilibrium/Gibbs/02}:
\begin{equation}\label{eq/Results/LocalEquilibrium/Table/02}
\begin{array}{rl}
\displaystyle \EuScript{E}_{\phi_{table}}(T, \psi) &= \displaystyle \sum_{i\,\in\, surface}{\left[\frac{\phi^{s}(\xs_{i}) - \phi^{s}_{eq}(\xs_{i}+x^{\ominus}_{eq}-x^{\ominus};\, T, \psi)}{\phi^{s}(\xs_{i})}\right]^{2}} \\
\\
\displaystyle \epsilon_{\phi_{table}}(\xs; T, \psi) &= \displaystyle \left| \frac{\phi^{s}(\xs)-\phi^{s}_{eq}(\xs; T, \psi)}{\phi^{s}(\xs)}\right|
\end{array}
\end{equation}
for each thermodynamic potential $h$, $u$, $f$, $g$. $x^{\ominus}$ and $x^{\ominus}_{eq}$ are the non-equilibrium and equilibrium positions of the reference surface in OFO. The set $\{\xs_{i}\}$
is the non-equilibrium surface grid and is used for both profiles. Since the width of an equilibrium surface may be not the same as the non-equilibrium one, the summation may exceed the formal
boundaries of the equilibrium surface. This is not a problem however, since the equilibrium profile $\psi^{s}_{eq}$ is the line with constant slope everywhere, as well as beyond the formal
boundaries. We don't shift the surface grid in the definition of $\epsilon_{\phi_{table}}(\xs; T, \psi)$ because in that notation $\xs$ means the particular dividing surface, while $\xs_{i}$
means the point of the surface grid.

The values of the corresponding errors are listed in \tblsr{tbl/Table/Tl}{tbl/Table/zil} in \ssecr{sec/ExcessErrors/Table}, and are found to be small.

As in \ssecr{sec/Results/LocalEquilibrium/Gibbs} we see, that for the equimolar surface the relative error in $g$ is huge. There is the same reason for this, namely that $g^{s}(x_{c}) \approx 0$
both in equilibrium and in non-equilibrium. This again makes the row corresponding to $g$ at $x^{c}$ be uninformative and one should not take into account these data.

\section{Conclusions}\label{sec/Conclusion}

The article continues the general analysis started in \cite{glav/grad1}. Here we focus on a specified mixture and develop further its properties. We choose a binary mixture of cyclohexane and
$n$-hexane as the system and describe in details how to implement the general analysis presented in \cite{glav/grad1}. We build numerical procedure for solving the resulted system of
differential equations. The resulted profiles of continuous variables are presented in \secr{sec/Results/Profile} and in the first article. We see, in particular, that a two component mixture
develop the temperature profile in the surface region which is similar to the temperature profile obtained for one-component system \cite{bedeaux/vdW/I}. Another characteristic of a binary
mixture is the difference between chemical potentials of components. The behavior of the profile of $\psi$ in non-equilibrium steady-states is different. It has different values in the different
bulk phases and we observe a transition from the one value to the other in the surface region.

We then proceed to verify the local equilibrium of the surface. This property means that a surface under non-equilibrium steady-state conditions can be described as equilibrium one in terms of
the Gibbs excess densities. We have discussed the meaning of the surface quantities in non-equilibrium and established the systematic procedure to obtain them. We were in particular focused on
i) the existence of the unique surface's temperature and chemical potential difference; ii) the validity of the relations between thermodynamic Gibbs excesses in a non-equilibrium surface; iii)
the correspondence between the non-equilibrium and equilibrium properties of the surface. It was possible to verify that one can speak about these statements independently on the choice of the
dividing surface. Similar results for the one-component system were obtained in \cite{bedeaux/vdW/II}.

The extrapolations procedure is numerical, and contains therefore a certain error. We may not expect this error to be negligible, not only because of numerical inaccuracy, but also because of
the non-equilibrium nature of the system. If the error is within a reasonable range, we will consider this as a satisfactory verification of local equilibrium.

The main part of the analysis in the interfacial region is introducing the excesses of thermodynamic densities, which are constructed with the help of extrapolated bulk profiles. In contrast to
equilibrium, non-equilibrium bulk profiles are not constants, and therefore their extrapolation to the surface region is not always accurate. The accuracy of extrapolation lowers when the
surface width increases. Apparent small deviations from local equilibrium are therefore to some extent an artefact of the inaccuracy of the extrapolation.

In the description of the surface excess densities it may happen that for particular choice of the dividing surface not one but several of the excesses are negligible. This increases the
relative error enormously while the absolute error remains finite and more or less constant. In order to avoid this problem we consider the excesses for all dividing surfaces together, rather
then for a particular dividing surface. Particularly, in \cite{bedeaux/vdW/II} the definition of the excess Gibbs energy was chosen differently because this excess was very small for the
equimolar surface. We have shown in this paper why this is not needed.

One can see from these data, that within different ways of perturbing mixture from equilibrium the biggest error comes when one perturbs the temperature on the liquid side. This is the most
extreme condition for the mixture being in non-equilibrium. While the relative temperature perturbation is only 2\%, the resulting temperature gradient is about $10^{8}$ K/m which is very far
beyond ordinary non-equilibrium conditions. The other perturbations make the validity of local equilibrium for the surface more precise. Similarly smaller perturbations make the validity of
local equilibrium also more precise.

We therefore conclude that the local equilibrium of the surface is valid with a reasonable accuracy.

\appendix

\section{Excesses' errors}\label{sec/ExcessErrors}
\subsection{Gibbs excesses' relative errors}\label{sec/ExcessErrors/Gibbs}
\begin{longtable}{l@{\qquad}l@{\qquad}l@{\qquad}l@{\qquad\qquad}l@{\qquad}l@{\qquad\qquad}}%
\caption{Gibbs excesses relative error for the case of perturbing $T^{\ell}$ in percent} \label{tbl/Gibbs/Tl} \\ %
\hline %
 &   &  \multicolumn{2}{c}{$T^{\ell} = 1.02\,T_{eq}$}  &  \multicolumn{2}{c}{$T^{\ell} = 0.98\,T_{eq}$}\\ %
\hline %
$\phi$ & error & for $T^{s}_{\xxs}, \psi^{s}_{\xxs}$ & for $T^{s}(\xs), \psi^{s}(\xs)$& for $T^{s}_{\xxs}, \psi^{s}_{\xxs}$ & for $T^{s}(\xs), \psi^{s}(\xs)$\\ %
\hline %
$h$ & $\EuScript{E}$ & 0.01328    & -& 0.033276   & - \\ %
 &  &  & &  & \\ %
  & $\epsilon_{\phi}(x^{c})$ & 0.023799   & 0.023419   & 0.064597   & 0.065002   \\ %
  & $\epsilon_{\phi}(x^{\gamma})$ & 0.020605   & 0.020737   & 0.056585   & 0.057148   \\ %
  & $\epsilon_{\phi}(x^{c_{1}})$ & 0.035878   & 0.033571   & 0.085212   & 0.085204   \\ %
  & $\epsilon_{\phi}(x^{c_{2}})$ & 0.015606   & 0.016541   & 0.047847   & 0.048578   \\ %
 \\ %
$u$ & $\EuScript{E}$ & 0.0071257  & -& 0.026714   & - \\ %
 &  &  & &  & \\ %
  & $\epsilon_{\phi}(x^{c})$ & 0.016729   & 0.016462   & 0.045109   & 0.045392   \\ %
  & $\epsilon_{\phi}(x^{\gamma})$ & 0.015059   & 0.015156   & 0.041048   & 0.041457   \\ %
  & $\epsilon_{\phi}(x^{c_{1}})$ & 0.022033   & 0.020616   & 0.054286   & 0.05428    \\ %
  & $\epsilon_{\phi}(x^{c_{2}})$ & 0.012161   & 0.012889   & 0.036244   & 0.036797   \\ %
 \\ %
$f$ & $\EuScript{E}$ & 0.20039    & -& 0.12983    & - \\ %
 &  &  & &  & \\ %
  & $\epsilon_{\phi}(x^{c})$ & 7.3727     & 7.3666     & 8.7959     & 8.799      \\ %
  & $\epsilon_{\phi}(x^{\gamma})$ & 1.9809     & 1.981      & 2.2605     & 2.2602     \\ %
  & $\epsilon_{\phi}(x^{c_{1}})$ & 1.6966     & 1.6881     & 2.5586     & 2.5602     \\ %
  & $\epsilon_{\phi}(x^{c_{2}})$ & 0.65429    & 0.65354    & 1.0109     & 1.0095     \\ %
 \\ %
$g$ & $\EuScript{E}$ & 0.36323    & -& 0.15487    & - \\ %
 &  &  & &  & \\ %
  & $\epsilon_{\phi}(x^{c})$ & 272.37     & 272.15     & 73.468     & 73.494     \\ %
  & $\epsilon_{\phi}(x^{\gamma})$ & 2.7241     & 2.7242     & 3.2313     & 3.2309     \\ %
  & $\epsilon_{\phi}(x^{c_{1}})$ & 1.4054     & 1.3983     & 1.9592     & 1.9605     \\ %
  & $\epsilon_{\phi}(x^{c_{2}})$ & 0.72857    & 0.72773    & 1.1818     & 1.1802     \\ %
 \\ %
\hline %
\end{longtable} %
\begin{longtable}{l@{\qquad}l@{\qquad}l@{\qquad}l@{\qquad\qquad}l@{\qquad}l@{\qquad\qquad}}%
\caption{Gibbs excesses relative error for the case of perturbing $p^{g}$ in percent} \label{tbl/Gibbs/pg} \\ %
\hline %
 &   &  \multicolumn{2}{c}{$p^{g} = 1.02\,p_{eq}$}  &  \multicolumn{2}{c}{$p^{g} = 0.98\,p_{eq}$}\\ %
\hline %
$\phi$ & error & for $T^{s}_{\xxs}, \psi^{s}_{\xxs}$ & for $T^{s}(\xs), \psi^{s}(\xs)$& for $T^{s}_{\xxs}, \psi^{s}_{\xxs}$ & for $T^{s}(\xs), \psi^{s}(\xs)$\\ %
\hline %
$h$ & $\EuScript{E}$ & 0.0013703  & -& 0.0019532  & - \\ %
 &  &  & &  & \\ %
  & $\epsilon_{\phi}(x^{c})$ & 0.0018964  & 0.0022352  & 0.0011722  & 0.00088793 \\ %
  & $\epsilon_{\phi}(x^{\gamma})$ & 0.0017063  & 0.0019978  & 0.0010683  & 0.00083823 \\ %
  & $\epsilon_{\phi}(x^{c_{1}})$ & 0.0025175  & 0.0030049  & 0.0015513  & 0.0010756  \\ %
  & $\epsilon_{\phi}(x^{c_{2}})$ & 0.0014259  & 0.0016451  & 0.00090372 & 0.00076269 \\ %
 \\ %
$u$ & $\EuScript{E}$ & 0.00041867 & -& 0.00033213 & - \\ %
 &  &  & &  & \\ %
  & $\epsilon_{\phi}(x^{c})$ & 0.0013316  & 0.0015695  & 0.00081921 & 0.00062057 \\ %
  & $\epsilon_{\phi}(x^{\gamma})$ & 0.00124    & 0.001452   & 0.00077388 & 0.00060722 \\ %
  & $\epsilon_{\phi}(x^{c_{1}})$ & 0.0015926  & 0.001901   & 0.00096161 & 0.00066669 \\ %
  & $\epsilon_{\phi}(x^{c_{2}})$ & 0.001093   & 0.001261   & 0.00069529 & 0.00058679 \\ %
 \\ %
$f$ & $\EuScript{E}$ & 0.043422   & -& 0.033692   & - \\ %
 &  &  & &  & \\ %
  & $\epsilon_{\phi}(x^{c})$ & 1.4194     & 1.4152     & 1.2845     & 1.2884     \\ %
  & $\epsilon_{\phi}(x^{\gamma})$ & 0.3884     & 0.38805    & 0.37919    & 0.3795     \\ %
  & $\epsilon_{\phi}(x^{c_{1}})$ & 0.40171    & 0.39739    & 0.30303    & 0.30621    \\ %
  & $\epsilon_{\phi}(x^{c_{2}})$ & 0.13128    & 0.1311     & 0.13976    & 0.13952    \\ %
 \\ %
$g$ & $\EuScript{E}$ & 0.08442    & -& 0.021892   & - \\ %
 &  &  & &  & \\ %
  & $\epsilon_{\phi}(x^{c})$ & 28.236     & 28.152     & 48.786     & 48.935     \\ %
  & $\epsilon_{\phi}(x^{\gamma})$ & 0.56258    & 0.56208    & 0.54645    & 0.5469     \\ %
  & $\epsilon_{\phi}(x^{c_{1}})$ & 0.3187     & 0.31528    & 0.24447    & 0.24704    \\ %
  & $\epsilon_{\phi}(x^{c_{2}})$ & 0.15021    & 0.15001    & 0.1583     & 0.15802    \\ %
 \\ %
\hline %
\end{longtable} %
\begin{longtable}{l@{\qquad}l@{\qquad}l@{\qquad}l@{\qquad\qquad}l@{\qquad}l@{\qquad\qquad}}%
\caption{Gibbs excesses relative error for the case of perturbing $\zeta^{\ell}$ in percent} \label{tbl/Gibbs/zil} \\ %
\hline %
 &   &  \multicolumn{2}{c}{$\zeta^{\ell} = 1.02\,\zeta^{\ell}_{eq}$}  &  \multicolumn{2}{c}{$\zeta^{\ell} = 0.98\,\zeta^{\ell}_{eq}$}\\ %
\hline %
$\phi$ & error & for $T^{s}_{\xxs}, \psi^{s}_{\xxs}$ & for $T^{s}(\xs), \psi^{s}(\xs)$& for $T^{s}_{\xxs}, \psi^{s}_{\xxs}$ & for $T^{s}(\xs), \psi^{s}(\xs)$\\ %
\hline %
$h$ & $\EuScript{E}$ & 0.00035435 & -& 0.0011482  & - \\ %
 &  &  & &  & \\ %
  & $\epsilon_{\phi}(x^{c})$ & 0.0023134  & 0.0028072  & 0.0042669  & 0.0042419  \\ %
  & $\epsilon_{\phi}(x^{\gamma})$ & 0.0021873  & 0.0026065  & 0.0039043  & 0.0039484  \\ %
  & $\epsilon_{\phi}(x^{c_{1}})$ & 0.0028054  & 0.0035457  & 0.0054811  & 0.0052479  \\ %
  & $\epsilon_{\phi}(x^{c_{2}})$ & 0.0020102  & 0.0023063  & 0.0033597  & 0.0035215  \\ %
 \\ %
$u$ & $\EuScript{E}$ & 0.00040969 & -& 0.0015073  & - \\ %
 &  &  & &  & \\ %
  & $\epsilon_{\phi}(x^{c})$ & 0.0016206  & 0.0019665  & 0.0029893  & 0.0029717  \\ %
  & $\epsilon_{\phi}(x^{\gamma})$ & 0.0015861  & 0.0018901  & 0.0028345  & 0.0028665  \\ %
  & $\epsilon_{\phi}(x^{c_{1}})$ & 0.0017517  & 0.0022139  & 0.0034446  & 0.003298   \\ %
  & $\epsilon_{\phi}(x^{c_{2}})$ & 0.0015408  & 0.0017677  & 0.0025842  & 0.0027086  \\ %
 \\ %
$f$ & $\EuScript{E}$ & 0.0063873  & -& 0.0026805  & - \\ %
 &  &  & &  & \\ %
  & $\epsilon_{\phi}(x^{c})$ & 0.17101    & 0.15419    & 0.033343   & 0.034999   \\ %
  & $\epsilon_{\phi}(x^{\gamma})$ & 0.014471   & 0.015093   & 0.023261   & 0.022844   \\ %
  & $\epsilon_{\phi}(x^{c_{1}})$ & 0.15718    & 0.13049    & 0.066572   & 0.055875   \\ %
  & $\epsilon_{\phi}(x^{c_{2}})$ & 0.062407   & 0.065842   & 0.037405   & 0.042978   \\ %
 \\ %
$g$ & $\EuScript{E}$ & 0.013595   & -& 0.0004864  & - \\ %
 &  &  & &  & \\ %
  & $\epsilon_{\phi}(x^{c})$ & 5.3413     & 4.816      & 0.77321    & 0.8116     \\ %
  & $\epsilon_{\phi}(x^{\gamma})$ & 0.021      & 0.021903   & 0.033471   & 0.032872   \\ %
  & $\epsilon_{\phi}(x^{c_{1}})$ & 0.12645    & 0.10498    & 0.052965   & 0.044454   \\ %
  & $\epsilon_{\phi}(x^{c_{2}})$ & 0.071253   & 0.075175   & 0.042467   & 0.048795   \\ %
 \\ %
\hline %
\end{longtable} %
\subsection{Equilibrium table excesses' relative errors}\label{sec/ExcessErrors/Table}
\begin{longtable}{l@{\qquad}l@{\qquad}l@{\qquad}l@{\qquad\qquad}l@{\qquad}l@{\qquad\qquad}}%
\caption{Equilibrium table excesses relative error for the case of perturbing $T^{\ell}$ in percent} \label{tbl/Table/Tl} \\ %
\hline %
 &   &  \multicolumn{2}{c}{$T^{\ell} = 1.02\,T_{eq}$}  &  \multicolumn{2}{c}{$T^{\ell} = 0.98\,T_{eq}$}\\ %
\hline %
$\phi$ & error & for $T^{s}_{\xxs}, \psi^{s}_{\xxs}$ & for $T^{s}(\xs), \psi^{s}(\xs)$& for $T^{s}_{\xxs}, \psi^{s}_{\xxs}$ & for $T^{s}(\xs), \psi^{s}(\xs)$\\ %
\hline %
$h$ & $\EuScript{E}$ & 0.67514    & -& 0.21243    & - \\ %
 &  &  & &  & \\ %
  & $\epsilon_{\phi}(x^{c})$ & 1.1052     & 1.1012     & 0.51202    & 0.50951    \\ %
  & $\epsilon_{\phi}(x^{\gamma})$ & 0.0062189  & 0.0050111  & 0.074177   & 0.071345   \\ %
  & $\epsilon_{\phi}(x^{c_{1}})$ & 8.5087     & 8.4753     & 6.4524     & 6.4581     \\ %
  & $\epsilon_{\phi}(x^{c_{2}})$ & 3.9073     & 3.8963     & 6.1946     & 6.1882     \\ %
 \\ %
$u$ & $\EuScript{E}$ & 0.32181    & -& 0.23381    & - \\ %
 &  &  & &  & \\ %
  & $\epsilon_{\phi}(x^{c})$ & 0.77685    & 0.77406    & 0.35744    & 0.3558     \\ %
  & $\epsilon_{\phi}(x^{\gamma})$ & 0.0043659  & 0.00366    & 0.053575   & 0.051756   \\ %
  & $\epsilon_{\phi}(x^{c_{1}})$ & 5.2257     & 5.2046     & 4.1104     & 4.1143     \\ %
  & $\epsilon_{\phi}(x^{c_{2}})$ & 3.0451     & 3.0361     & 4.692      & 4.6875     \\ %
 \\ %
$f$ & $\EuScript{E}$ & 0.0046133  & -& 0.0035049  & - \\ %
 &  &  & &  & \\ %
  & $\epsilon_{\phi}(x^{c})$ & 6.6737     & 6.67       & 8.0558     & 8.0599     \\ %
  & $\epsilon_{\phi}(x^{\gamma})$ & 7.242      & 7.2428     & 0.37022    & 0.37236    \\ %
  & $\epsilon_{\phi}(x^{c_{1}})$ & 15.74      & 15.7       & 25.631     & 25.642     \\ %
  & $\epsilon_{\phi}(x^{c_{2}})$ & 13.328     & 13.306     & 20.092     & 20.077     \\ %
 \\ %
$g$ & $\EuScript{E}$ & 0.0017615  & -& 0.001069   & - \\ %
 &  &  & &  & \\ %
  & $\epsilon_{\phi}(x^{c})$ & 246.55     & 246.41     & 67.29      & 67.32      \\ %
  & $\epsilon_{\phi}(x^{\gamma})$ & 9.9591     & 9.9604     & 0.52885    & 0.53228    \\ %
  & $\epsilon_{\phi}(x^{c_{1}})$ & 13.038     & 13.006     & 19.627     & 19.635     \\ %
  & $\epsilon_{\phi}(x^{c_{2}})$ & 14.84      & 14.817     & 23.49      & 23.472     \\ %
 \\ %
\hline %
\end{longtable} %
\begin{longtable}{l@{\qquad}l@{\qquad}l@{\qquad}l@{\qquad\qquad}l@{\qquad}l@{\qquad\qquad}}%
\caption{Equilibrium table excesses relative error for the case of perturbing $p^{g}$ in percent} \label{tbl/Table/pg} \\ %
\hline %
 &   &  \multicolumn{2}{c}{$p^{g} = 1.02\,p_{eq}$}  &  \multicolumn{2}{c}{$p^{g} = 0.98\,p_{eq}$}\\ %
\hline %
$\phi$ & error & for $T^{s}_{\xxs}, \psi^{s}_{\xxs}$ & for $T^{s}(\xs), \psi^{s}(\xs)$& for $T^{s}_{\xxs}, \psi^{s}_{\xxs}$ & for $T^{s}(\xs), \psi^{s}(\xs)$\\ %
\hline %
$h$ & $\EuScript{E}$ & 0.26501    & -& 1.3628     & - \\ %
 &  &  & &  & \\ %
  & $\epsilon_{\phi}(x^{c})$ & 0.79783    & 0.80083    & 0.90685    & 0.90956    \\ %
  & $\epsilon_{\phi}(x^{\gamma})$ & 1.4549     & 1.457      & 1.2481     & 1.2505     \\ %
  & $\epsilon_{\phi}(x^{c_{1}})$ & 1.3056     & 1.2901     & 2.0582     & 2.0702     \\ %
  & $\epsilon_{\phi}(x^{c_{2}})$ & 2.3955     & 2.3962     & 0.069749   & 0.072757   \\ %
 \\ %
$u$ & $\EuScript{E}$ & 0.047434   & -& 0.20486    & - \\ %
 &  &  & &  & \\ %
  & $\epsilon_{\phi}(x^{c})$ & 0.5599     & 0.56232    & 0.63362    & 0.63569    \\ %
  & $\epsilon_{\phi}(x^{\gamma})$ & 1.0571     & 1.0589     & 0.90402    & 0.90589    \\ %
  & $\epsilon_{\phi}(x^{c_{1}})$ & 0.82642    & 0.81614    & 1.2756     & 1.2832     \\ %
  & $\epsilon_{\phi}(x^{c_{2}})$ & 1.836      & 1.8368     & 0.053565   & 0.055977   \\ %
 \\ %
$f$ & $\EuScript{E}$ & 0.003912   & -& 0.0035128  & - \\ %
 &  &  & &  & \\ %
  & $\epsilon_{\phi}(x^{c})$ & 1.218      & 1.2144     & 1.1886     & 1.1919     \\ %
  & $\epsilon_{\phi}(x^{\gamma})$ & 4.7735     & 4.775      & 2.1767     & 2.1788     \\ %
  & $\epsilon_{\phi}(x^{c_{1}})$ & 7.4201     & 7.3951     & 2.2105     & 2.2271     \\ %
  & $\epsilon_{\phi}(x^{c_{2}})$ & 5.5423     & 5.5439     & 1.9531     & 1.9467     \\ %
 \\ %
$g$ & $\EuScript{E}$ & 0.0013182  & -& 0.00041813 & - \\ %
 &  &  & &  & \\ %
  & $\epsilon_{\phi}(x^{c})$ & 24.251     & 24.157     & 45.122     & 45.27      \\ %
  & $\epsilon_{\phi}(x^{\gamma})$ & 6.9147     & 6.9164     & 3.1371     & 3.1398     \\ %
  & $\epsilon_{\phi}(x^{c_{1}})$ & 5.8866     & 5.8671     & 1.7835     & 1.7967     \\ %
  & $\epsilon_{\phi}(x^{c_{2}})$ & 6.3419     & 6.3437     & 2.212      & 2.2049     \\ %
 \\ %
\hline %
\end{longtable} %
\begin{longtable}{l@{\qquad}l@{\qquad}l@{\qquad}l@{\qquad\qquad}l@{\qquad}l@{\qquad\qquad}}%
\caption{Equilibrium table excesses relative error for the case of perturbing $\zeta^{\ell}$ in percent} \label{tbl/Table/zil} \\ %
\hline %
 &   &  \multicolumn{2}{c}{$\zeta^{\ell} = 1.02\,\zeta^{\ell}_{eq}$}  &  \multicolumn{2}{c}{$\zeta^{\ell} = 0.98\,\zeta^{\ell}_{eq}$}\\ %
\hline %
$\phi$ & error & for $T^{s}_{\xxs}, \psi^{s}_{\xxs}$ & for $T^{s}(\xs), \psi^{s}(\xs)$& for $T^{s}_{\xxs}, \psi^{s}_{\xxs}$ & for $T^{s}(\xs), \psi^{s}(\xs)$\\ %
\hline %
$h$ & $\EuScript{E}$ & 0.093953   & -& 0.20248    & - \\ %
 &  &  & &  & \\ %
  & $\epsilon_{\phi}(x^{c})$ & 0.85205    & 0.86417    & 0.83511    & 0.8363     \\ %
  & $\epsilon_{\phi}(x^{\gamma})$ & 1.5557     & 1.5598     & 1.1532     & 1.1601     \\ %
  & $\epsilon_{\phi}(x^{c_{1}})$ & 0.24334    & 0.33902    & 0.26902    & 0.23261    \\ %
  & $\epsilon_{\phi}(x^{c_{2}})$ & 1.265      & 1.2405     & 1.2654     & 1.3065     \\ %
 \\ %
$u$ & $\EuScript{E}$ & 0.068203   & -& 0.24097    & - \\ %
 &  &  & &  & \\ %
  & $\epsilon_{\phi}(x^{c})$ & 0.59647    & 0.60537    & 0.585      & 0.58586    \\ %
  & $\epsilon_{\phi}(x^{\gamma})$ & 1.1278     & 1.1311     & 0.8372     & 0.84222    \\ %
  & $\epsilon_{\phi}(x^{c_{1}})$ & 0.15127    & 0.21168    & 0.169      & 0.14618    \\ %
  & $\epsilon_{\phi}(x^{c_{2}})$ & 0.96943    & 0.95081    & 0.97333    & 1.0049     \\ %
 \\ %
$f$ & $\EuScript{E}$ & 0.0026249  & -& 0.0035651  & - \\ %
 &  &  & &  & \\ %
  & $\epsilon_{\phi}(x^{c})$ & 0.095422   & 0.080499   & 0.18152    & 0.18276    \\ %
  & $\epsilon_{\phi}(x^{\gamma})$ & 4.6443     & 4.6454     & 2.382      & 2.3867     \\ %
  & $\epsilon_{\phi}(x^{c_{1}})$ & 2.3591     & 2.2252     & 2.7907     & 2.8466     \\ %
  & $\epsilon_{\phi}(x^{c_{2}})$ & 1.6382     & 1.5842     & 1.897      & 1.9858     \\ %
 \\ %
$g$ & $\EuScript{E}$ & 0.00098739 & -& 0.0010871  & - \\ %
 &  &  & &  & \\ %
  & $\epsilon_{\phi}(x^{c})$ & 3.025      & 2.5143     & 4.2062     & 4.237      \\ %
  & $\epsilon_{\phi}(x^{\gamma})$ & 6.7404     & 6.7414     & 3.4276     & 3.4343     \\ %
  & $\epsilon_{\phi}(x^{c_{1}})$ & 1.8975     & 1.7901     & 2.2203     & 2.2648     \\ %
  & $\epsilon_{\phi}(x^{c_{2}})$ & 1.8705     & 1.8088     & 2.1538     & 2.2546     \\ %
 \\ %
\hline %
\end{longtable} %
%

%

\bibliographystyle{unsrt}

\end{document}